\documentclass[twocolumn,prd,noshowpacs,aps,a4paper,10pt,nofootinbib]{revtex4}

\usepackage{bm}
\usepackage{amsmath}
\usepackage{amssymb}

\usepackage{graphicx}
\usepackage{hyperref}
\usepackage{color}

\usepackage{geometry}
\newcommand{\tr}[1]{\mathrm{Tr}\left[#1\right]}

\usepackage{enumitem}
\DeclareMathAlphabet\mathbfcal{OMS}{cmsy}{b}{n}
\usepackage[utf8]{inputenc}
\usepackage{braket}
\usepackage{scrextend}
\deffootnote{0em}{1.6em}{\thefootnotemark.\enskip}

\begin{document}

%--------------------------------------------------------------------------------------------------------------------------------------------------------------------------
%HEADER
%--------------------------------------------------------------------------------------------------------------------------------------------------------------------------
\title{Quadratic estimator for CMB cross-correlation}

\author{S. Vanneste}
\email{vanneste@lal.in2p3.fr}
\author{S. Henrot-Versillé}
\author{T. Louis}
\author{M. Tristram}
\affiliation{LAL, Université Paris-Sud, CNRS/IN2P3, Université Paris-Saclay, Orsay, France}

\begin{abstract}
The Quadratic Maximum Likelihood estimator can be used to reconstruct the Cosmic Microwave Background (CMB) power spectra with minimal error bars. Still, it requires an accurate estimate of the datasets noise covariance matrix in order to be corrected for spurious bias. We describe an extension of this method to cross-correlation, thus removing noise bias and mitigating the impact of systematic effects, providing that they are uncorrelated. This estimator is tested on two simulation surveys at large and intermediate angular scales, respectively corresponding to satellite and ground-based CMB experiments. The analysis focuses on polarization maps, over a wide range of noise levels from $0.1$ to $50 \, \mu \rm K.arcmin$. We show how this estimator minimizes the increase of variance due to polarization leakage between $E$ and $B$ modes. We compare this method with the pure pseudo-spectrum formalism which is computationally faster but less optimal, especially on large angular scales.
\end{abstract}

\maketitle

%---------------------------------------------------------------------------------------------------------------------------------
%INTRODUCTION
%---------------------------------------------------------------------------------------------------------------------------------
\section{Introduction}

The precise characterization of the Comic Microwave Background (CMB) polarization will provide a wealth of information in addition to the {\it Planck} satellite CMB temperature measurement~\cite{planck_collaboration_planck_2016}. It will help to further constrain the $\Lambda \rm CDM$ cosmological model and its extensions. CMB polarization is generally described in terms of the two linear components Stokes parameters $Q$ and $U$, which can be mathematically combined to define the curl-free '$E$' and divergence-free '$B$' polarization patterns. CMB anisotropies are conveniently projected in harmonic space, with their statistics encoded in the angular power spectra $C_\ell^{XY}$, where $\ell$ is the multipole, and $X,Y \in \{E,B\}$. Since the anisotropies in the CMB are expected to be Gaussian distributed, all the cosmological information is contained in $C_\ell$.

\begin{figure}[!htb]
\includegraphics[width=\columnwidth]{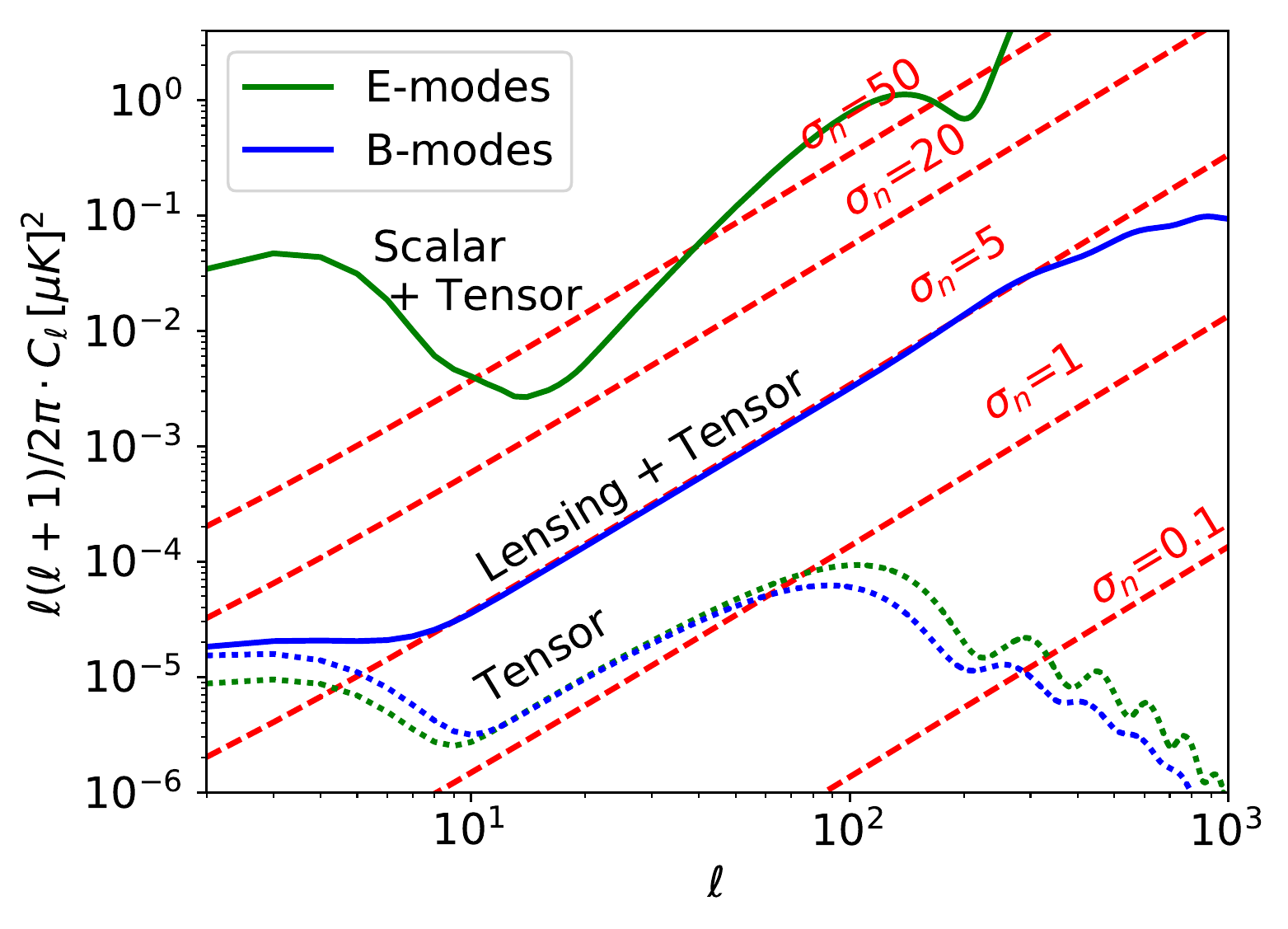}
\caption{Tensor (dashed) and total (solid) components of the $E$-modes (green), and $B$-modes (blue) spectra $\ell(\ell+1)/(2\pi) \cdot C_\ell$ as a function of the multipole $\ell$, based on {\it Planck} 2015 best fit model with an optical depth $\tau=0.06$. The primordial (tensor) polarization spectra are indicated for a tensor-to-scalar ratio $r=10^{-3}$. Various experimental noise levels $\sigma_n\,[\mu\rm K . arcmin]$ are indicated.\label{fig:ClthVSmuK}}
\end{figure}

The dominant source of $E$-modes anisotropies are scalar fluctuations at the epoch of recombination. Tensor (primordial gravitational waves) perturbations generated during inflation can act as a subdominant source of $E$-modes. Primordial $B$-modes, however, are only sourced by tensor fluctuations, and thus represent a unique observable to test inflationary physics. Their amplitude, parametrized by the tensor-to-scalar ratio '$r$', can be arbitrarily low, depending on the inflation energy scale, and is expected to be maximal at large and intermediate angular scales ($\ell\lesssim 10^2$). In addition, CMB photons undergo a lensing effect induced by their passage through the gravitational field of matter between the CMB last scattering surface and us, which leads to the transfer of $E$ to $B$ modes (and vice versa). The lensing $B$-modes thus contaminate the primordial $B$-modes signal. Figure~\ref{fig:ClthVSmuK} represents the $E$-modes and the predicted lensing + tensor $B$-modes derived from the {\it Planck} best fit model~\cite{planck_collaboration_planck_2016} with an optical depth parameter $\tau=0.06$~\cite{planck_collaboration_planck_2016-1}. In addition, $E$ and $B$ tensor modes are shown for $r=10^{-3}$, as well as instrumental noise levels between $0.1$ and $50\,\mu\rm K.arcmin$.

Because of experimental limitations and/or foreground contaminations, the effective CMB surveys sky coverage can be partial. This introduces an ambiguity in the relationship between the Stokes parameters and the $E$ and $B$ modes. In this context, the $E$ and $B$ modes are inevitably mixed and mislabeled~\cite{bunn_separating_2003,bunn_e/b_2003}. Although this polarization leakage can be corrected on average~\cite{chon_fast_2004,kogut_firstyear_2003}, the $E/B$ mixing signals contribute to each other's spectrum variance. Since the $B$-mode signal is expected to be much lower than the $E$-mode signal, the impact of this 'variance leakage' is extremely problematic for the detection of $B$-modes and their precise measurement.

The pure pseudo-spectrum (PpCl) method presented in~\cite{smith_pseudo-$c_ell$_2006,bunn_separating_2003,lewis_harmonic_2003} is an extension of the standard pseudo-spectrum method (pCl) and currently represents the most popular solution that reduces the amount of polarization variance leakage. It has been widely investigated in e.g. \cite{grain_polarized_2009,grain_cmb_2012}, and has been demonstrated to produce near-optimal variance power spectrum estimates for intermediate and small angular scales. The extension of the PpCl method to cross-spectra formalism offers the advantage of cross-correlate CMB maps, allowing us to remove correlated noise and mitigate the impact of systematic effects, providing that they are uncorrelated. However, the PpCl method requires particular sky mask apodizations, which depend on the scanning strategy and on the depth of the observed CMB field. Moreover, the method has been proved to be sub-optimal for large and intermediate angular scales ($\ell \lesssim 100$)~\cite{molinari_comparison_2014,efstathiou_myths_2004,efstathiou_hybrid_2006}.

Other methods consist of estimating the spectra using a pixel based approach, which is particularly relevant for large angular scale analysis, but they have the drawback of being computationally more expensive. The Maximum Likelihood Estimator (MLE) and the Quadratic Maximum Likelihood (QML) have the advantage of minimizing spectra uncertainties. The latter, developed in~\cite{tegmark_how_1997} and extended to polarization in~\cite{tegmark_how_2001}, gives the same error bars as the MLE and requires $\mathcal O (N_{\rm d}^3)$ operations for a dataset of size $N_{\rm d}$, relative to the pCl which only demands $\mathcal O (N_{\rm d}^{3/2})$ operations~\cite{efstathiou_myths_2004}. 

In this paper we describe a method based on the QML approach that allows us to cross-correlate CMB maps that have common sky coverage, in analogy with the pseudo cross-spectra formalism. The formalism was first introduced in \cite{planck_collaboration_planck_2016_XLVI} for the 2016 Planck results.
Although this spectrum estimator is not derived from a maximum likelihood, we will refer to it as the cross-QML (xQML) for simplicity. The analysis presented hereafter focuses on the case of polarization spectra and the xQML ability to reduce the impact of $E/B$ variances leakage. 

In Sec.~\ref{sec:method}, we develop the formalism of the xQML estimator. We review the QML and extend it to cross-spectra. We then discuss its bias and uncertainty. Important steps of the xQML implementation are then described, in particular the pixel covariance matrix construction, and the binning of the spectrum estimator.
In Sec.~\ref{sec:MCSimu}, the xQML is tested on two simulation set-ups : a large angular scale survey aiming at the measurement of the reionization signal ($\ell \lesssim 10$) , and an intermediate angular scale survey aiming at the measurement of the recombination bump ($\ell \simeq 100$). We show that the xQML method is unbiased, and gives minimum error bars.
The polarization leakage is discussed in Sec.~\ref{sec:Leakage}, in which we also compare the xQML $B$-mode variance with other methods such as the PpCl.
The same analysis is realized in Sec.~\ref{sec:EBspec} for the $EB$ spectrum.
We conclude in Sec.~\ref{sec:Conclusion}, and we forecast the uncertainty on $r$ based on the different methods introduced and compared in this paper.

%---------------------------------------------------------------------------------------------------------------------------------
% METHOD
%---------------------------------------------------------------------------------------------------------------------------------
\section{\label{sec:method}Method}

In this section we review the most important steps that lead to the definition of the QML estimator, following what has been done in \cite{tegmark_how_1997,tegmark_how_2001}. We then derive a cross-spectrum QML estimator (xQML) and compare its properties with the QML. Finally, we discuss in depth the implementation of the algorithm.

Lower case characters correspond to vectors and upper case correspond to matrices. Bold font, Latin indices, the trace and transpose operators are used for elements in the pixel domain, while normal font and $\ell$ indices are used in the multipole domain.

We consider a dataset $\mathbf{d}$, of dimension $N_d = 3 n_\text{pix}$ which encodes temperature and Stokes parameters measurements,
\begin{equation}
\mathbf{d} \equiv \begin{pmatrix}\mathbf T \\ \mathbf Q \\ \mathbf U  \end{pmatrix}.
\end{equation}

The pixels covariance matrix $\mathbf C$ of the dataset is given by 
\begin{equation}
\mathbf C \equiv \braket{\mathbf{d},{\mathbf{d}}^T} = \mathbf S + \mathbf N, \label{eq:PixelCov}
\end{equation}
with $\mathbf N$ the pixel noise covariance matrix, and $\mathbf S$ the signal covariance matrix defined as
\begin{equation}
\mathbf S \equiv \sum_{\ell} \mathbf P_\ell C_\ell, \quad {\rm with} \quad \mathbf P_\ell^{ij} = \frac{\partial \mathbf C^{ij}}{\partial C_\ell}. \label{eq:Smatrix} 
\end{equation}
The vector $C_\ell$ encodes all six power spectra $TT, EE, BB, TE, TB,$ and $EB$.

\subsection{\label{sec:QML}QML estimator}

We review important steps of the QML estimator developed in~\cite{tegmark_how_1997,tegmark_how_2001}. We can write the power spectrum estimator as a quadratic function of the pixels
\begin{equation}
\hat y_\ell \equiv \mathbf d^T \mathbf E_\ell \mathbf d - b_\ell. \label{eq:autoyl}
\end{equation}
$\mathbf E_\ell$ ($\ell=2,...$) are arbitrary $N_d \times N_d$ matrices, and $b_\ell$ are arbitrary constants.
From Eqs. \eqref{eq:PixelCov} and \eqref{eq:Smatrix}, the estimator ensemble average reads
\begin{flalign}
\braket{\hat y_\ell} & = \tr{ \mathbf E_\ell \braket{\mathbf d, \mathbf d^T}} -  b_\ell , \\
& =\sum_{\ell'}  W_{\ell \ell'} C_{\ell'} + \tr { \mathbf E_\ell \mathbf N}  -  b_\ell,
\end{flalign}
with 
\begin{equation}
W_{\ell\ell'} \equiv\tr{\mathbf E_\ell \mathbf P_{\ell'}} \label{eq:autoWindow}
\end{equation}
as the 'mode-mixing' matrix. Choosing $b_\ell = \tr { \mathbf E_\ell \mathbf N}$, the unbiased estimator of the true power spectrum $C_\ell$ thus reads
\begin{equation}
\hat C_\ell \equiv \sum_{\ell'}  [W^{-1}]_{\ell\ell'} \hat y_{\ell'}, \label{eq:autoCl}
\end{equation}
and has the following covariance
\begin{equation}
\braket{\Delta \hat C_\ell,  \Delta \hat C_{\ell'}} =   [W^{-1}]_{\ell\ell_1} \braket{\Delta  \hat y_{\ell_1}, \Delta  \hat y_{\ell_2}} [W^{-1}]_{\ell_2\ell'}, \label{eq:autoCovCl}
\end{equation}
where $\Delta \hat C_\ell = \hat C_\ell - \braket{\hat C_\ell}$. The summation over repeated indices is implied. The resulting power spectrum is unbiased, regardless of the choice of the $\mathbf E_\ell$ matrices. However, they are usually constructed in order to minimize the estimator variance
\begin{equation}
\braket{\Delta \hat y_\ell,  \Delta \hat y_{\ell}} = 2 \tr{\mathbf C \mathbf E_\ell \mathbf C \mathbf E_{\ell}}, \label{eq:autoCovyl}
\end{equation}
which gives the trivial solution $\mathbf E_\ell = \mathbf 0$. We thus impose the mode-mixing matrix diagonal to be nonzero. For each $\ell$, introducing the Lagrange multipliers $\lambda$ and the condition $W_{\ell \ell} = \beta$, we require the derivative of
\begin{equation}
\braket{\Delta \hat y_\ell,  \Delta \hat y_{\ell}} -  2 \lambda ( \tr{\mathbf E_\ell \mathbf P_{\ell}} - \beta), \label{eq:autoLagrange}
\end{equation}
with respect to $\mathbf E_\ell$ to vanish, and obtain the solution\footnote{Using matrix identities $\partial_{\mathbf E} \tr{\mathbf C\mathbf E\mathbf C\mathbf E} = 2 \mathbf C^T \mathbf E^T \mathbf C^T $.}
\begin{equation}
\mathbf E_\ell = \frac\lambda2  \mathbf C^{-1} \mathbf P_{\ell} \mathbf C^{-1}. \label{eq:autoEl}
\end{equation}
Finally, imposing $W_{\ell \ell}= \tr{\mathbf E_\ell \mathbf P_{\ell}} = \beta$ gives
\begin{equation}
\lambda  \tr{ \mathbf C^{-1} \mathbf P_{\ell} \mathbf C^{-1} \mathbf P_{\ell}} = \beta .
\end{equation}
We choose $\beta$ such that $\lambda= 1$ and $\mathbf E_\ell$ is well defined. With this choice, the mode-mixing matrix $W_{\ell\ell'}$  is the Fisher information matrix 
\begin{equation}
F_{\ell\ell'} = \frac12 \tr{  \mathbf C^{-1} \mathbf P_{\ell} \mathbf C^{-1} \mathbf P_{\ell'}}, \label{eq:autoFisher}
\end{equation}
with $\braket{ \Delta  \hat y_\ell,  \Delta \hat y_{\ell'}} = F_{\ell\ell'}$ and $\braket{\Delta \hat C_\ell,  \Delta \hat C_{\ell'}} = [F^{-1}]_{\ell\ell'}$.

The $\mathbf E_\ell$ matrices are thus constructed such that the spectrum estimator has minimal variance. However, the QML estimator requires a precise knowledge of the pixel noise matrix $\mathbf N$ to cancel the bias term $b_\ell$ in Eq.~\eqref{eq:autoyl}. In practice, estimating the noise model of an experiment is difficult and requires an exquisite knowledge of instrument properties. In the next section, we develop a method that allows us to compute a cross-spectrum estimator that is unbiased independently of the choice of $\mathbf N$.

\subsection{\label{sec:xQMLmethod}xQML estimator}

Following the same formalism as for the 'auto'-spectrum QML estimator detailed in the previous section, we now consider two datasets $\mathbf{d}^A$ and $\mathbf{d}^B$ from which the pixel covariance matrix reads 
\begin{equation}
\mathbf C^{AB}\equiv\braket{\mathbf{d}^A,{\mathbf{d}^B}^T}=\mathbf S + \mathbf N^{AB}.	
\end{equation}
We assume uncorrelated noise between the two datasets, such that the cross pixel noise covariance matrix vanishes $\mathbf N^{AB}=0$.

The cross estimator now reads
\begin{equation}
\hat y_\ell^{AB} \equiv {\mathbf d^A}^T \mathbf E_\ell \mathbf d^B - b_\ell^{AB}, \label{eq:crossyl}
\end{equation}
with $b_\ell^{AB} = \tr { \mathbf E_\ell \mathbf N^{AB}} = 0$.

The covariance of the estimator is computed using Wick's theorem,
\begin{flalign}
&\braket{\Delta \hat y^{AB}_\ell, \Delta \hat y^{AB}_{\ell'}} \nonumber\\
&= \left[ \braket{d^A_i,d^A_k}\braket{d^B_j,d^B_n}\right. + \left.\braket{d^A_i,d^B_n}\braket{d^A_j,d^B_k} \right] \mathbf E_\ell^{ij}\mathbf E_{\ell'}^{kn}\nonumber \\
&= \tr{\mathbf C^{AA} \mathbf E_\ell \mathbf C^{BB} \mathbf E^T_{\ell'} + \mathbf C^{AB} \mathbf E_\ell \mathbf C^{AB} \mathbf E_{\ell'}}, \label{eq:crossCovyl}
\end{flalign}
where summation on the pixels indices $i,j,k,n$ is implied. Matrices $\mathbf C^{AA} = \mathbf S + \mathbf N^{AA}$ and $\mathbf C^{BB} = \mathbf S + \mathbf N^{BB}$ are respectively the pixel covariance matrix of the datasets $A$ and $B$. As in Eqs.~\eqref{eq:autoCl} and \eqref{eq:autoCovCl} for the QML method, the unbiased estimator reads
\begin{equation}
\hat C_\ell \equiv \sum_{\ell'} [W^{-1}]_{\ell\ell'} \hat y^{AB}_{\ell'}, \label{eq:crossCl}
\end{equation}
and its covariance
\begin{equation} 
\braket{\Delta \hat C_{\ell},  \Delta \hat C_{\ell'}} = [W^{-1}]_{\ell\ell_1} \braket{\Delta \hat y^{AB}_{\ell_1},  \Delta \hat y^{AB}_{\ell_2}} [W^{-1}]_{\ell_2\ell'}.\label{eq:crossCovCl}
\end{equation}

As in Eq.~\eqref{eq:autoLagrange} for the QML, we seek for the $\mathbf E_\ell$ matrices that minimize the estimator variance of Eq.~\eqref{eq:crossCovyl}. We get the equation\footnote{Using matrix identities $\partial_{\mathbf E} \tr{\mathbf A\mathbf E\mathbf B\mathbf E} = \mathbf A^T \mathbf E^T \mathbf B^T + \mathbf B^T \mathbf E^T \mathbf A^T$ and $\partial_{\mathbf E} \tr{\mathbf A\mathbf E\mathbf B \mathbf E^T} = \mathbf A^T \mathbf E \mathbf B^T + \mathbf A \mathbf E  \mathbf B$.}
\begin{eqnarray}
&  \mathbf C^{AA} \mathbf E_\ell \mathbf C^{BB} + \mathbf C^{AB} \mathbf E_\ell^T \mathbf C^{AB}  = \lambda\mathbf P_{\ell}, \label{eq:Sylvester}
\end{eqnarray}
which is a generalized form of the Sylvester equation~\cite{de_teran_uniqueness_2016}. Although the exact solution exists, as discussed in Sec.~\ref{sec:Implementation}, it requires us to solve a system of $N_d^2$ equations, which quickly becomes computationally prohibitive for large datasets. For this reason, we derive an approximate solution by considering two extreme signal-to-noise ratio (SNR) cases~:
\setlist[description]{font=\normalfont\itshape\space}
\begin{description}
\item[\textsc{Hs}] : High SNR, such that \\ 
$\mathbf S \gg \mathbf N$, and $\mathbf C^{AA} \sim \mathbf C^{BB} \sim \mathbf S$.
\item[\textsc{Ls}] : Low SNR , such that \\ 
$\mathbf S \ll \mathbf N$, and $\mathbf C^{AA} \sim \mathbf N^{AA} $, $\mathbf C^{BB} \sim \mathbf N^{BB}$.
\end{description}

For both limits, Eq~\eqref{eq:Sylvester} admits a solution of the form\footnote{We remark that when $\mathbf C^{AA} \sim \mathbf C^{BB}$, and more specifically for a high signal-to-noise ratio, $\mathbf E_\ell \simeq \mathbf E_\ell^T$.}
\begin{eqnarray}
\mathbf E_\ell \simeq \frac \lambda \alpha  (\mathbf C^{AA})^{-1} \mathbf P_{\ell} (\mathbf C^{BB})^{-1}, \label{eq:crossEl}
\end{eqnarray}
where $ \alpha$ is a normalization coefficient that depends on the SNR, with $\alpha=2$ for the \textsc{Hs} regime, and $\alpha=1$ for the \textsc{Ls} regime. The impact of the approximation made in Eq.~\eqref{eq:crossEl} on the spectrum variance is discussed in Sec~\ref{sec:Implementation}. Finally, imposing $ W_{\ell \ell} = \tr{\mathbf E_\ell \mathbf P_{\ell}} = \beta$ gives
\begin{equation}
\frac\lambda\alpha  \tr{ (\mathbf C^{AA})^{-1} \mathbf P_{\ell} (\mathbf C^{BB})^{-1} \mathbf P_{\ell}} = \beta.
\end{equation}
We choose $\beta$ such that $\lambda/\alpha = 1/2$, and we recover the QML solution for $A=B$. Inserting $\mathbf E_\ell$ of Eq.~\eqref{eq:crossEl} in the mode-mixing matrix defined in Eq.~\eqref{eq:autoWindow}, one obtains
\begin{equation}
W_{\ell\ell'} = \frac 12 \tr{(\mathbf C^{AA})^{-1} \mathbf P_{\ell} (\mathbf C^{BB})^{-1} \mathbf P_{\ell'} }.\label{eq:crossFisher}
\end{equation}
Using Eqs.~\eqref{eq:crossCovyl},~\eqref{eq:crossCovCl}, \eqref{eq:crossEl} and~\eqref{eq:crossFisher}, the cross-spectrum estimator covariance reads
\begin{flalign}
\braket{\Delta \hat C_\ell,  \Delta \hat C_{\ell'}} &= \frac12  [W^{-1}]_{\ell \ell_1}  \left( W_{\ell_1 \ell_2} +G _{\ell_1 \ell_2} \right) [W^{-1}]_{\ell_2 \ell'} \nonumber \\
& = \frac12 \left ([ W^{-1}] _{\ell \ell'} + [ W^{-1}] _{\ell \ell_1}  G _{\ell_1 \ell_2} [ W^{-1}] _{\ell_2 \ell'}\right) \nonumber \\
&\equiv V_{\ell \ell'}, \label{eq:crossCovClEl}
\end{flalign}
where we define
\begin{equation}
\begin{split}
G_{\ell \ell'} \equiv \frac12 \mathrm{Tr} \left[ (\mathbf C^{AA})^{-1} \mathbf P_{\ell} (\mathbf C^{BB})^{-1}  \mathbf C^{AB}   \right.  \\  \times \left. (\mathbf C^{AA})^{-1} \mathbf P_{\ell'} (\mathbf C^{BB})^{-1}  \mathbf C^{AB} \right]. \label{eq:crossGisher}
\end{split}
\end{equation}

\begin{figure}[!htb]
\includegraphics[width=\columnwidth]{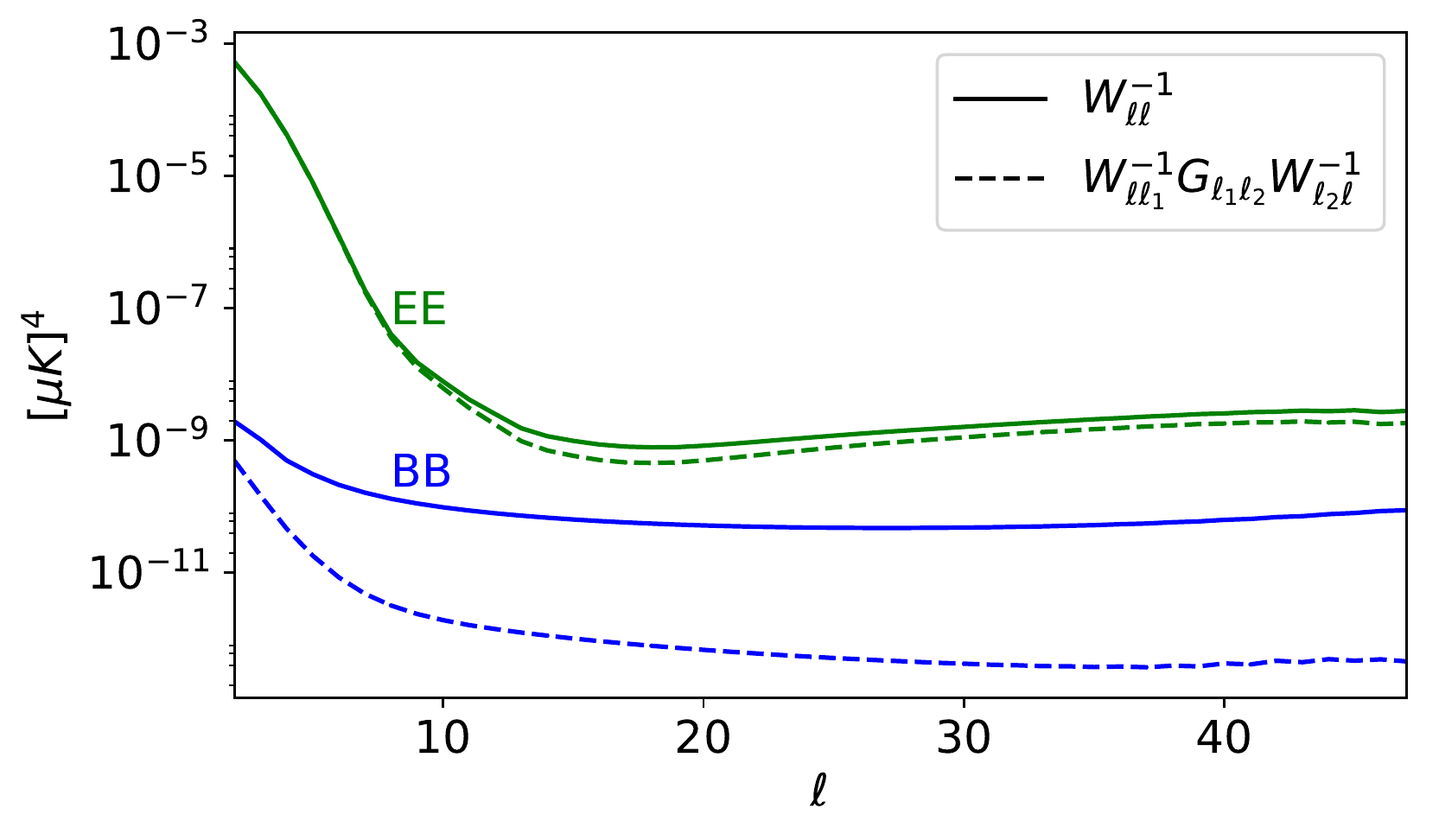}
\caption{Diagonals of the covariance matrix terms $W_{\ell \ell_1}^{-1} G _{\ell_1 \ell_2} W_{\ell_2 \ell}^{-1}$ (dashed) and $W_{\ell \ell}^{-1}$ (plain) of Eq.~\eqref{eq:crossCovClEl}. $EE$ and $BB$ components are plotted in green and blue respectively. The noise level is $10\,\mu{\rm K . arcmin}$.\label{fig:FvsG}}
\end{figure}

In the \textsc{Hs} regime $G_{\ell\ell'} \sim W_{\ell \ell'}$, such that $V_{\ell\ell'} = [ W^{-1}] _{\ell \ell'}$. In the \textsc{Ls} regime, the second term $[W^{-1}] _{\ell \ell_1}  G _{\ell_1 \ell_2} [ W^{-1}] _{\ell_2 \ell'}$ in Eq.~\eqref{eq:crossCovClEl} contributes at second order of the cross-spectrum variance. As a representative example, the diagonal elements of those two terms are compared in Fig.~\ref{fig:FvsG} for the $EE$ and $BB$ spectra, with a $10\,\mu\rm K.arcmin$ noise level. With this choice, the $E$-mode is signal dominated, and corresponds to the \textsc{Hs} regime, while the $B$-mode SNR is low for most of the multipoles ($\ell \gtrsim10$), and corresponds to the \textsc{Ls} case.

We successfully defined a quadratic estimator based on datasets cross-correlation which does not require the subtraction of noise bias. Moreover, we derived an approximation of the $\mathbf E_\ell$ matrices that minimizes its variance. We also recover the QML estimator when $A=B$, with a nonvanishing noise bias term $b_\ell$. 

\subsection{\label{sec:Implementation}Implementation}

In this section we detail some important steps of the xQML implementation. We first discuss the pixel covariance matrix construction. We then derive an exact solution for the Sylvester Eq.~\eqref{eq:Sylvester}. Finally, we describe a method for binning the xQML spectrum estimator.

\subsubsection{Pixel covariance matrix}

The $\mathbf P_\ell$ matrices Eq.~\eqref{eq:Smatrix} are directly multiplied by each of the datasets pixel window and beams transfer functions. We do not discuss the matrices construction, for which further details can be found in~\cite{tegmark_how_2001}.

The covariance matrix $\mathbf C$ introduced in Eq.~\eqref{eq:PixelCov} includes correlations between pixels for each of the Stokes parameters,
\begin{eqnarray} 
\mathbf C = \begin{pmatrix} C^{TT} & C^{TQ} & C^{TU} \\ C^{QT} & C^{QQ} & C^{QU} \\ C^{UT} & C^{UQ} & C^{UU}\end{pmatrix}. \label{eq:Cstokes}
\end{eqnarray}

We can separate the temperature and polarization measurements by using an approximated pixel covariance matrix
\begin{eqnarray} 
 \tilde {\mathbf C} = \begin{pmatrix} C^{TT} & 0 & 0\\ 0 & C^{QQ} & C^{QU} \\ 0 & C^{UQ} & C^{UU}\end{pmatrix} .
\end{eqnarray}

This matrix does not mix temperature with polarization estimates. As a result, the $\hat C_\ell$ estimator is not optimal anymore, while it is still an unbiased estimator of the true $C_\ell$. As shown in~\cite{tegmark_how_2001}, the price to pay is a slight error bar increase of the order of one percent. Using this choice, temperature and polarization analysis can be done completely separately. For the rest of this paper, we focus our analysis on polarization measurement only. The method can be implemented for the temperature spectrum estimation following the same approach.

In Eq.~\eqref{eq:Smatrix}, the summation over $\ell$ is theoretically infinite. It can however be truncated at a given $\ell_{max}$ as long as the remaining contributions from $C_{\ell > \ell_{max}}$ are negligible. This can be accomplished manually by smoothing the dataset $\mathbf d$ (e.g. by convolving the spectrum with a decreasing function). In the framework of our analysis, we simply generated CMB simulations while filtering all $C_{\ell > \ell_{max}}$.

The xQML variance has been shown to be minimal if the fiducial $\tilde {\mathbf C}$ matrix is built from the true ${\mathbf C}$. In practice, it is not always possible to estimate precisely the latter. We can compute the estimator variance in Eq.~\eqref{eq:crossCovCl} for any fiducial  $\tilde {\mathbf C}$
\begin{equation}
\begin{split}
&\braket{\Delta \hat C_\ell, \Delta  \hat C_{\ell'}} \equiv \\
&[\tilde W^{-1}]_{\ell \ell_1} \tr{\mathbf C^{AA} \tilde{\mathbf E}_{\ell_1} \mathbf C^{BB} \tilde{\mathbf E}^T_{\ell_2}}  [\tilde W^{-1}]_{\ell_2 \ell'} \label{eq:CovClmc} \\
&+ [\tilde W^{-1}]_{\ell \ell_1} \tr{\mathbf C^{AB} \tilde{\mathbf E}_{\ell_1} \mathbf C^{AB} \tilde{\mathbf E}_{\ell_2}} [\tilde W^{-1}]_{\ell_2 \ell'},
\end{split}
\end{equation}
where $\tilde{\mathbf E}_\ell$ and $\tilde{\mathbf W}_{\ell\ell'}$ are computed using $\tilde {\mathbf C}$ in Eqs.~\eqref{eq:crossEl} and \eqref{eq:crossFisher}. 

To estimate the impact on the spectra estimations variance of small deviations of the fiducial $\tilde {\mathbf C}$ from the true ${\mathbf C}$, we consider a simplified toy model with $\mathbf C^{AA} = \mathbf C^{BB} = \mathbf C$. We also restrict our calculation to the first term of Eq.~\eqref{eq:CovClmc}, since we showed that, depending on the noise level, the second term is either negligible, or either equal to the first one. Any small perturbation to the fiducial $\tilde {\mathbf C}$ around the true  $\mathbf C$ can be written as 
\begin{equation}
\tilde {\mathbf C} = \mathbf C + \mathbfcal E, \quad {\rm with }\quad \mathbfcal E \ll \mathbf C,
\end{equation}
and thus 
\begin{equation}
\tilde {\mathbf C}^{-1} = \mathbf C^{-1} - \mathbfcal D, \; {\rm with }\; \mathbfcal D \equiv \mathbf C^{-1}\mathbfcal E\mathbf C^{-1} \ll \mathbf C^{-1}
\end{equation}
 At the first order in $\mathbfcal D$,
\begin{equation}
\tr{\mathbf C \tilde{\mathbf E}_{\ell} \mathbf C \tilde{\mathbf E}_{\ell}}  \simeq \tr{\mathbf C^{-1} \mathbf P_\ell \mathbf C^{-1} \mathbf P_{\ell} - 4 \mathbf C^{-1} \mathbf P_\ell \mathbfcal D \mathbf{P_{\ell}}},
\end{equation}
and
\begin{equation}
\tilde{ W}_{\ell\ell} \simeq \tr{\mathbf C^{-1} \mathbf P_\ell \mathbf C^{-1} \mathbf P_{\ell} - 2 \mathbf C^{-1} \mathbf P_\ell \mathbfcal D \mathbf{P_{\ell}} }.
\end{equation}
Inserting both expressions in Eq.~\eqref{eq:CovClmc},
\begin{flalign}
&\braket{\Delta \hat C_\ell, \Delta  \hat C_{\ell'}} = \nonumber\\
&\frac{\tr{\mathbf C^{-1} \mathbf P_\ell \mathbf C^{-1} \mathbf P_{\ell}} - 4\tr{\mathbf C^{-1} \mathbf P_\ell \mathbfcal D \mathbf{P_{\ell}}} }{ \left(\tr{\mathbf C^{-1} \mathbf P_\ell \mathbf C^{-1} \mathbf P_{\ell} } - 2 \tr{\mathbf C^{-1} \mathbf P_\ell \mathbfcal D \mathbf{P_{\ell}}}\right)^2 } \\
& \simeq   \tr{\mathbf C^{-1} \mathbf P_\ell \mathbf C^{-1} \mathbf P_{\ell}}^{-1} \\
& = V_{\ell \ell}.
\end{flalign}

We see that a fiducial $\tilde {\mathbf C}$ sufficiently close to the true ${\mathbf C}$ induces only second order deviations of the spectrum estimation variance from the optimal variance $V_{\ell\ell'}$. For a low SNR, the choice of the fiducial $\tilde C_\ell$ have little impact on $\tilde {\mathbf C}$. Conversely, for signal dominated datasets, deviations of $\tilde C_\ell$ can have a non-negligible impact on the spectrum error. A solution is to run the xQML method iteratively as recommended in \cite{tegmark_how_2001}, with previous spectrum estimation as the new fiducial model. This especially applies for the tensor-to-scalar ratio and reionization fiducial parameters. We have found that the choice of their fiducial values, if far from their true values, can greatly increase the large angular scale uncertainty of $BB$ and $EE$ spectra.

However, even if the variance of the spectrum estimation is only slightly impacted when the fiducial $\tilde {\mathbf C}$ diverges from the true dataset covariance matrix, the analytical estimate of the variance $\tilde V_{\ell\ell'} = \dfrac12 \left ([ \tilde W^{-1}] _{\ell \ell'} + [ \tilde W^{-1}] _{\ell \ell_1}  \tilde G _{\ell_1 \ell_2} [ \tilde W^{-1}] _{\ell_2 \ell'}\right) $ is biased. Taking, for example, $\tilde {\mathbf C} = \gamma \mathbf C $ implies $\braket{\Delta \hat C_\ell, \Delta  \hat C_{\ell'}} = V_{\ell\ell'}$, but $ \tilde V_{\ell\ell'} =  \gamma ^2 V_{\ell\ell'} $, for any constant $\gamma$. One must thus be cautious when estimating the spectrum variance analytically.

\subsubsection{\label{sec:ElApprox}Sylvester equation solution}

We discuss the approximate solution of Eq.~\eqref{eq:Sylvester} introduced in Sec.~\ref{sec:xQMLmethod}, also known as a generalized form of the Sylvester equation, and we compare it with the exact solution described in~\cite{de_teran_uniqueness_2016}. To find the exact solution, we use the Kronecker product property ${\rm vec}(\mathbf A \mathbf X \mathbf B) = (\mathbf B^T \otimes \mathbf A) {\rm vec}(\mathbf X)$, under the condition that the product $\mathbf A \mathbf X \mathbf B$ is well defined. The operator ${\rm vec}()$ vectorizes a matrix (by stacking its columns), and $\otimes$ is the Kronecker matrix product. We also introduce the permutation matrix $\Pi$ such that ${\rm vec} (\mathbf X^T)=\Pi\,{\rm vec} (\mathbf X)$. One can show that ${\rm vec}(\mathbf A \mathbf X^T \mathbf B) = \Pi\,(\mathbf A^T \otimes \mathbf B) {\rm vec}(\mathbf X)$~\cite{horn_topics_1991}. The Sylvester Eq.~\eqref{eq:Sylvester} can thus be written as a set of linear equations
\begin{equation}
\begin{split}
\left[\mathbf C^{BB} \otimes \mathbf C^{AA} + \Pi (\mathbf C^{AB}\otimes \mathbf C^{AB})\right]{\rm vec}(\mathbf E_\ell)  \\
= {\rm vec}(\mathbf P_{\ell}). \label{eq:vecSylvester}
\end{split}{}
\end{equation}
We can then solve it exactly for ${\rm vec}(\mathbf E_\ell)$ using the least-squares method. However, the equation system is of dimension $N_d^2$, which quickly becomes computationally costly for large datasets. Using Eq.~\eqref{eq:CovClmc} with the approximate solution in Eq.~\eqref{eq:crossEl} as $\tilde{\mathbf E}_\ell$, we find that the deviation of the spectrum variance from the minimum one using the exact solution of Eq.~\eqref{eq:vecSylvester} is of the order of $2\%$ in the worse case, when the signal and the noise level are of the same order. We can thus safely use the approximated solution of Eq.~\eqref{eq:crossEl} for the implementation of the xQML method.

\subsubsection{\label{sec:Binning}Binning}{}

CMB observations are only available on a limited sky fraction, and as a result, individual multipoles can be strongly correlated when reconstructing the CMB spectra. It is thus convenient to bin the power spectra in multipoles band powers, labeled $b$ hereafter. We define the binning operators,
\begin{equation}
R_{b\ell} = \begin{cases} \Delta_b^{-1} \\ 0 \end{cases}, Q_{\ell b} = \begin{cases} 1  \quad  \text{ if } \ell\in b  \\ 0 \quad \text{ otherwise } \end{cases}\label{eq:binoperator},
\end{equation}
with $\Delta_{b}$ the width of the $b$th bin, which can be varied from one bin to another. The binned estimator is written 
\begin{equation}
\hat y_b \equiv \sum_{\ell} R_{b \ell} \hat y_\ell, \label{eq:binyl}
\end{equation}
for which the covariance reads
\begin{flalign}
\braket{\Delta \hat y_b, \Delta y_{b'}} = & \tr{\mathbf C^{AA} \mathbf E_b \mathbf C^{BB} \mathbf E^T_{b'}+\mathbf C^{AB} \mathbf E_b \mathbf C^{AB} \mathbf E_{b'}} \nonumber\\
= & \frac 1{2 \Delta_{b'}} \left ( W_{b b'}+ G _{b b'}\right) \label{eq:CovylBin}
\end{flalign}
with $ W_{b b'} = R_{b\ell} W_{\ell \ell'} Q_{\ell' b'}$, and $ G_{b b'} = R_{b\ell} G_{\ell \ell'} Q_{\ell' b'}$. The true binned spectrum is thus 
\begin{equation}
C_b \equiv \sum_{\ell,\ell',b'}  [W^{-1}]_{b b'} R_{b'\ell} W_{\ell \ell'} C_{\ell'},
\end{equation}
and its unbiased binned estimation becomes 
\begin{equation}
\hat C_b \equiv  \sum_{\ell'} [W^{-1}]_{b b'} \hat y_{b'}, \label{eq:binCl}
\end{equation} 
with covariance
\begin{equation}
V_{bb'} =\frac 1{2 \Delta_{b'}} \left ( [W^{-1}]_{bb'}+[W^{-1}]_{bb_1} G _{b_1 b_2}  [W^{-1}]_{b_2b'} \right). \label{eq:binCovCl}
\end{equation}

We remark that the binning can also be achieved by computing $\mathbf P_{b} \equiv \sum_{\ell\in b} \mathbf P_{\ell}$ directly (without the normalization term $\Delta_{b}$), or equivalently $\mathbf P_{b} \equiv  \sum_{\ell} \mathbf P_{\ell} Q_{\ell b}$. With this definition of $\mathbf P_{b}$, the xQML components can be computed as usually defined in Eqs.~\eqref{eq:crossyl}, \eqref{eq:crossCl}, \eqref{eq:crossEl} and \eqref{eq:crossFisher} for the spectrum estimate $\hat C_\ell$, and Eqs.~\eqref{eq:crossCovClEl}, \eqref{eq:crossGisher} and \eqref{eq:crossFisher} for its analytical covariance (replacing all subscripts $\ell$ by $b$). This method is computationally more efficient compared to the method presented above.

%--------------------------------------------------------------------------------------------------------------------------------------------------------------------------
%SIMULATIONS
%--------------------------------------------------------------------------------------------------------------------------------------------------------------------------
\begin{figure*}[htb]
\includegraphics[width=0.49\textwidth]{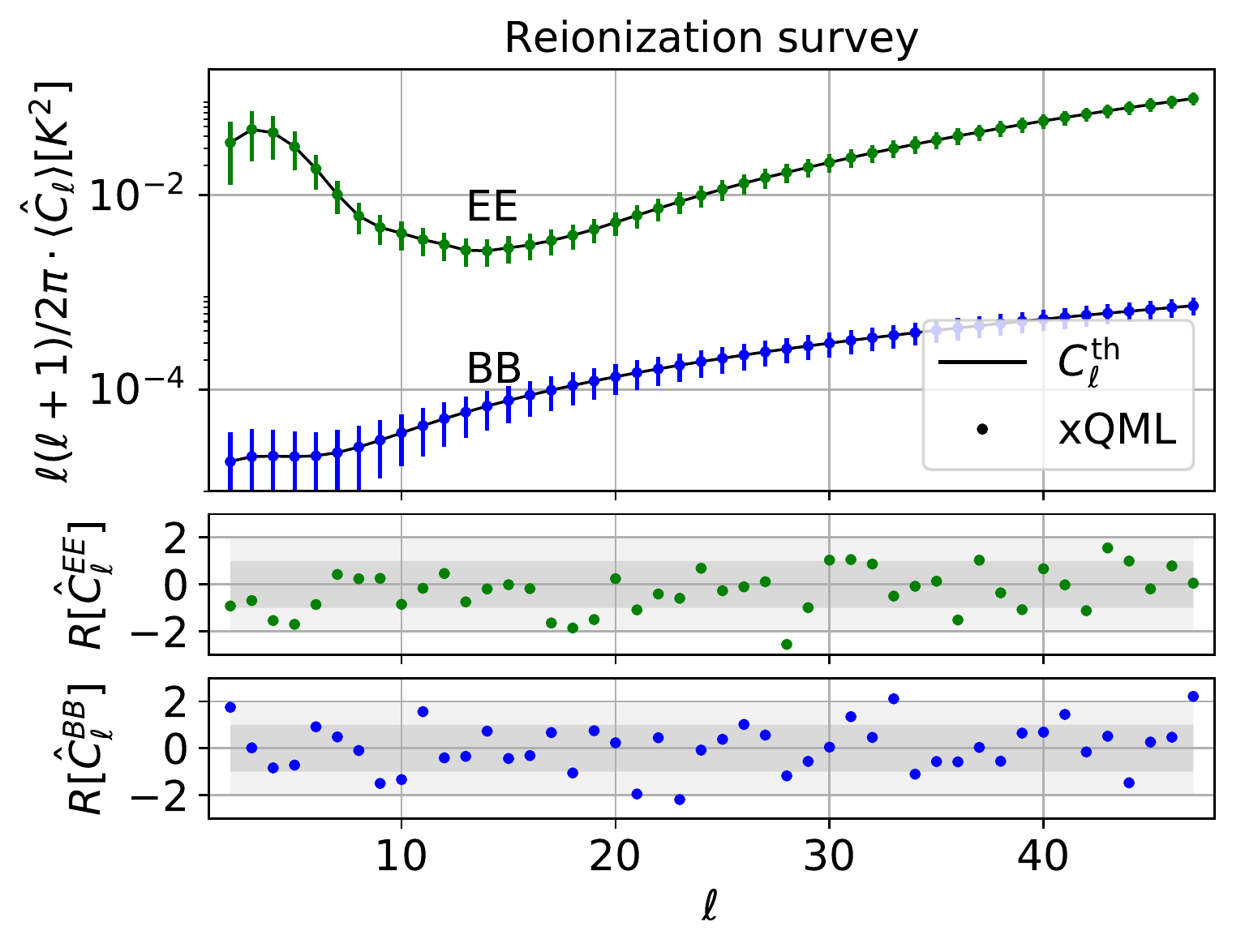}
\includegraphics[width=0.49\textwidth]{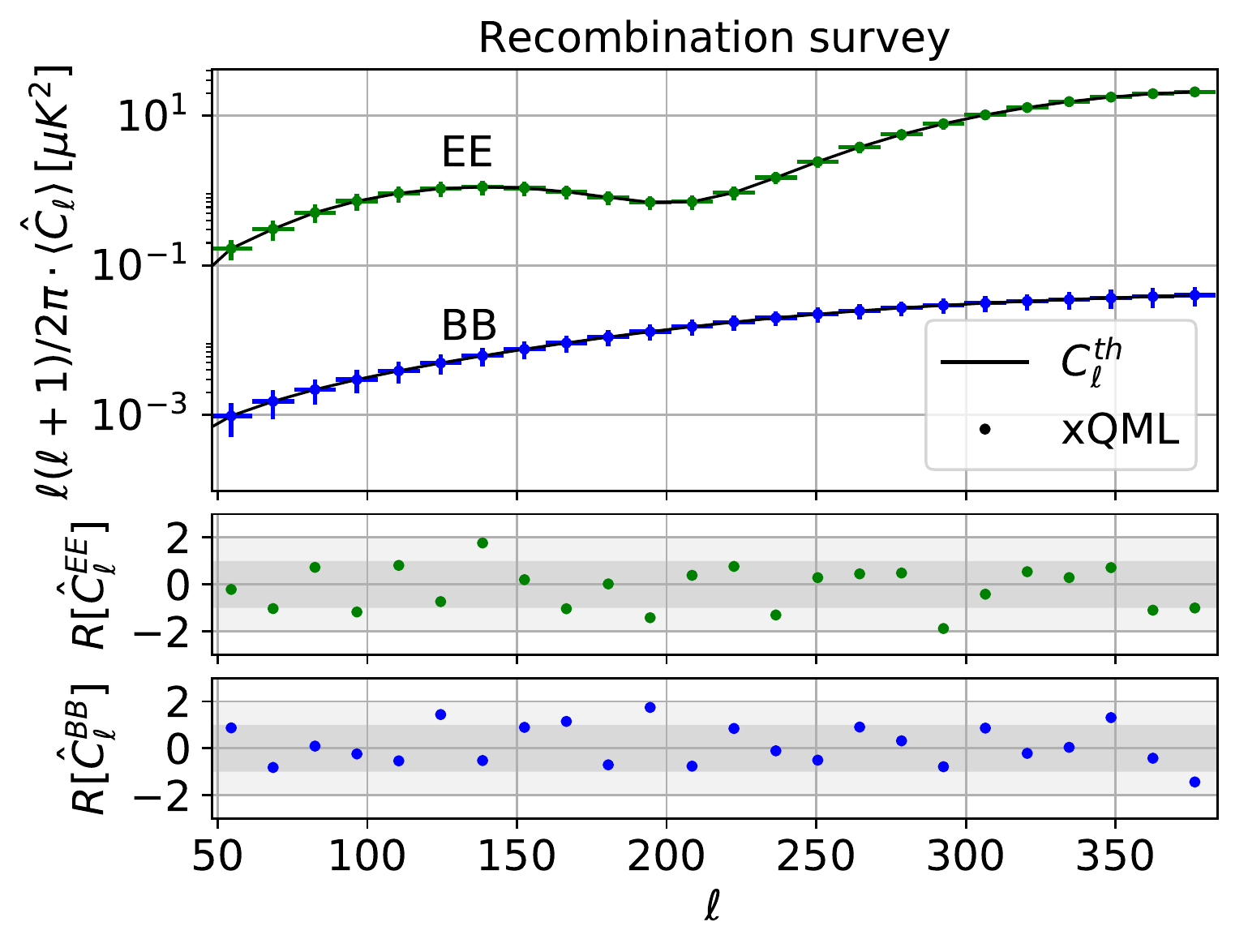}
\caption{$EE$ (green) and $BB$ (blue) mean power spectra xQML estimates $\ell(\ell+1)/2\pi \cdot \braket{\hat C_\ell}$, and residues $R_\ell[\hat C_\ell]$ from Eq.~\eqref{eq:ResidSpect}, computed from $n_{\rm MC}=10^5$ Monte-Carlo (MC) simulations. Spectra models are plotted in black solid lines. Left panel corresponds to the reionization survey simulations ($n_{side} = 16$, $f_{\rm sky}\simeq 0.7\%$), right panel corresponds to the recombination survey simulations ($n_{side} = 128$, $f_{\rm sky}\simeq 1\%$). Noise level is $\sigma_n = 1\,\mu \rm K.arcmin$ for both surveys.\label{fig:ThvsMC}}
\end{figure*}

%---------------------------------------------------------------------------------------------------------------------------------
% Monte Carlo
%---------------------------------------------------------------------------------------------------------------------------------

\section{\label{sec:MCSimu}Monte Carlo Simulations}

\begin{figure}[!htb]
\includegraphics[width=\columnwidth]{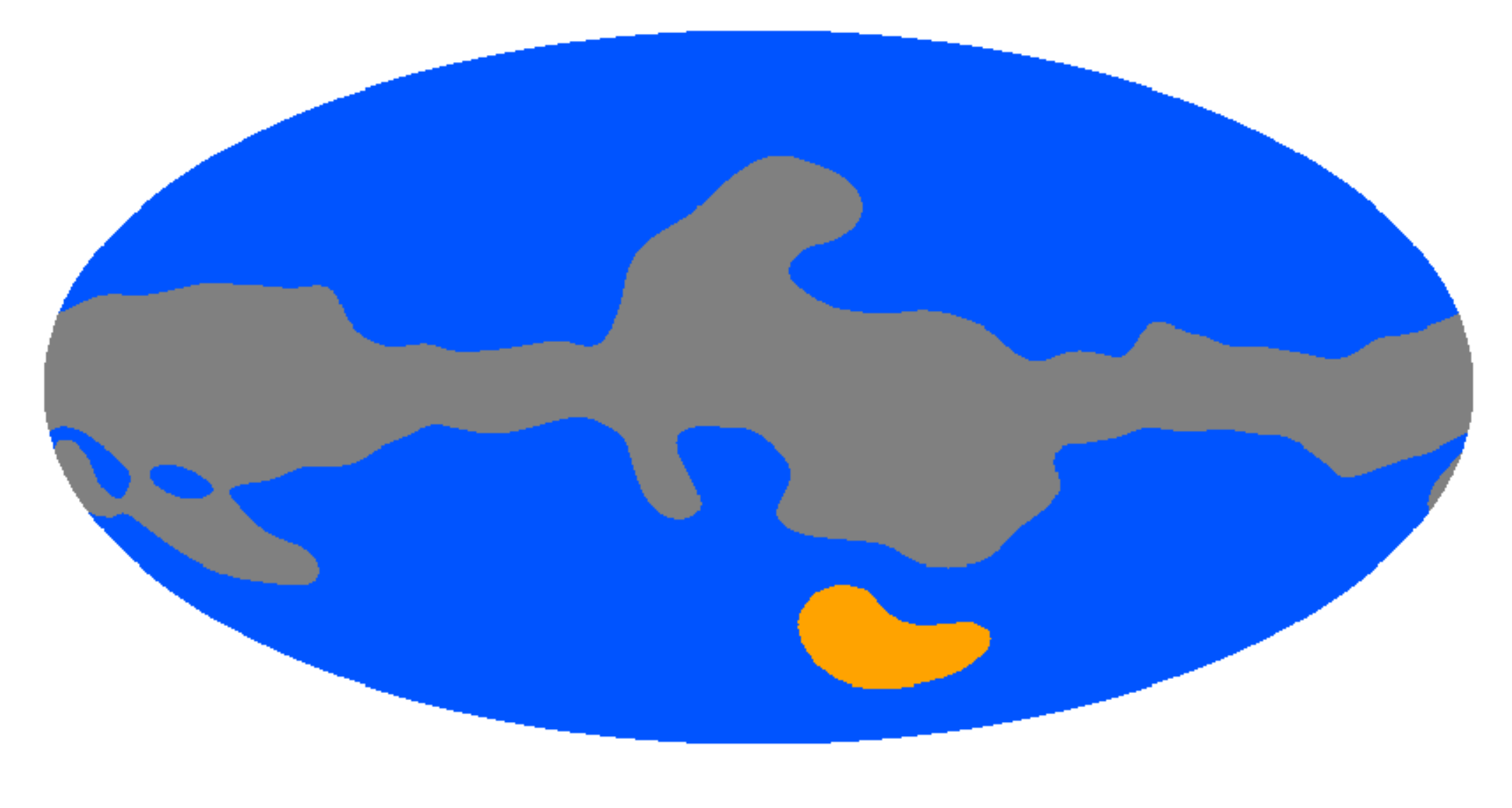}
\caption{Mollweide projection of the sky coverages for the reionization (yellow + blue areas), and the recombination (yellow area) surveys. The latter corresponds to the $\sim1\%$ sky fraction from {\it BICEP2} public mask. The grey area corresponds to the $30\%$ where {\it Planck} dust polarization amplitude is the highest, mostly located in the galactic plane.\label{fig:Masks}}
\end{figure}

In this section we describe two simulated surveys on which we test the xQML estimator. We first consider a full sky experiment aiming at the measurement of the reionization signal ($\ell \lesssim 10$). The foreground contaminations are assumed to be removed, and their residuals, which are assumed to be strong in the galactic plane, are masked. The second survey covers a smaller sky fraction, aiming at the measurement of the recombination bump ($\ell \simeq 100$), and for which the foregrounds contamination is assumed to be removed. Both surveys sky fractions are shown in Fig.~\ref{fig:Masks}. We generate $n_{MC} = 10^5$ CMB simulations from the {\it Planck} 2015 best fit spectrum model~\cite{planck_collaboration_planck_2016} shown in Fig.~\ref{fig:ClthVSmuK}, with a tensor-to-scalar ratio $r=10^{-3}$, and a reionization optical depth $\tau=0.06$. The two surveys are treated completely independently. For each of them, we cross-correlated two simulated maps (usually obtained through data-splits), with noise levels between $0.1 \leq \sigma_n \leq 50 \,\mu\rm K.arcmin$ indicated in Fig.~\ref{fig:ClthVSmuK}. This choice roughly covers the characteristics of future ground experiments from CMB Stage 4 (S4)~\cite{abazajian_cmb-s4_2016} ($\sim 1\,\mu\rm K.arcmin$), or satellites such as {\it LiteBIRD}, {\it CORE}, and {\it PICO} (between $1$ and $5\,\mu\rm K.arcmin$)~\cite{matsumura_mission_2014,delabrouille_exploring_2018,de_zotti_prospects_2018}, up to {\it Planck} noise level (around $50\,\mu\rm K.arcmin$)~\cite{the_planck_collaboration_scientific_2006}.

\subsection{Reionization survey}

For the large angular scales analysis, referred as the 'reionization survey', we consider an observed sky fraction $f_{\rm sky} \simeq 70\%$. A binary mask is built from the $353\,\rm GHz$ {\it Planck} polarization maps, for which pixels with the highest polarization amplitude $(Q^2+U^2)^{1/2}$ accurately traces the galactic polarized dust. We choose to follow the instrumental specifications of the satellite mission {\it LiteBIRD}~\cite{matsumura_mission_2014}, considering a beam-width of $0.5\,\rm deg$, and a white homogeneous noise. The analysis is done at the map resolution {$ n_{side} = 16$}, over the multipoles range $\ell\in[2,47]$.

\subsection{Recombination survey}

The 'recombination survey' sky patch is based on the public {\it BICEP2}~\cite{noauthor_keck_nodate} apodized mask $M \in [0,1]$. We build a binary mask using all pixels $i$ for which $M_i \geq 0.1$. Rather than considering a homogeneous noise as for the reionization survey, we apply an inverse weighting noise distribution based on the mask $M$. The effective sky fraction is therefore $f_{\rm sky} = (\sum_i{M_i^2})^2/\sum_i M_i^4 \simeq 1\%$, as defined in~\cite{hivon_master_2002}. Our analysis is done with maps resolution {$ n_{side} = 128$}, and a beam-width of $0.5\,\rm deg$. Because of the limited sky fraction, individual multipoles are strongly correlated. We thus reconstruct the spectrum using the binning scheme described in Sec.~\ref{sec:Implementation}. We show the results starting from $\ell=48$ to account for the insensitivity of the survey to large angular scales, and we define $24$ bins up to {$\ell=383$} with $\Delta_{b} = 14$.

%--------------------------------------------------------------------------------------------------------------------------------------------------------------------------
%RESULTS
%--------------------------------------------------------------------------------------------------------------------------------------------------------------------------

\begin{figure*}[!hbt]
\includegraphics[width=\columnwidth]{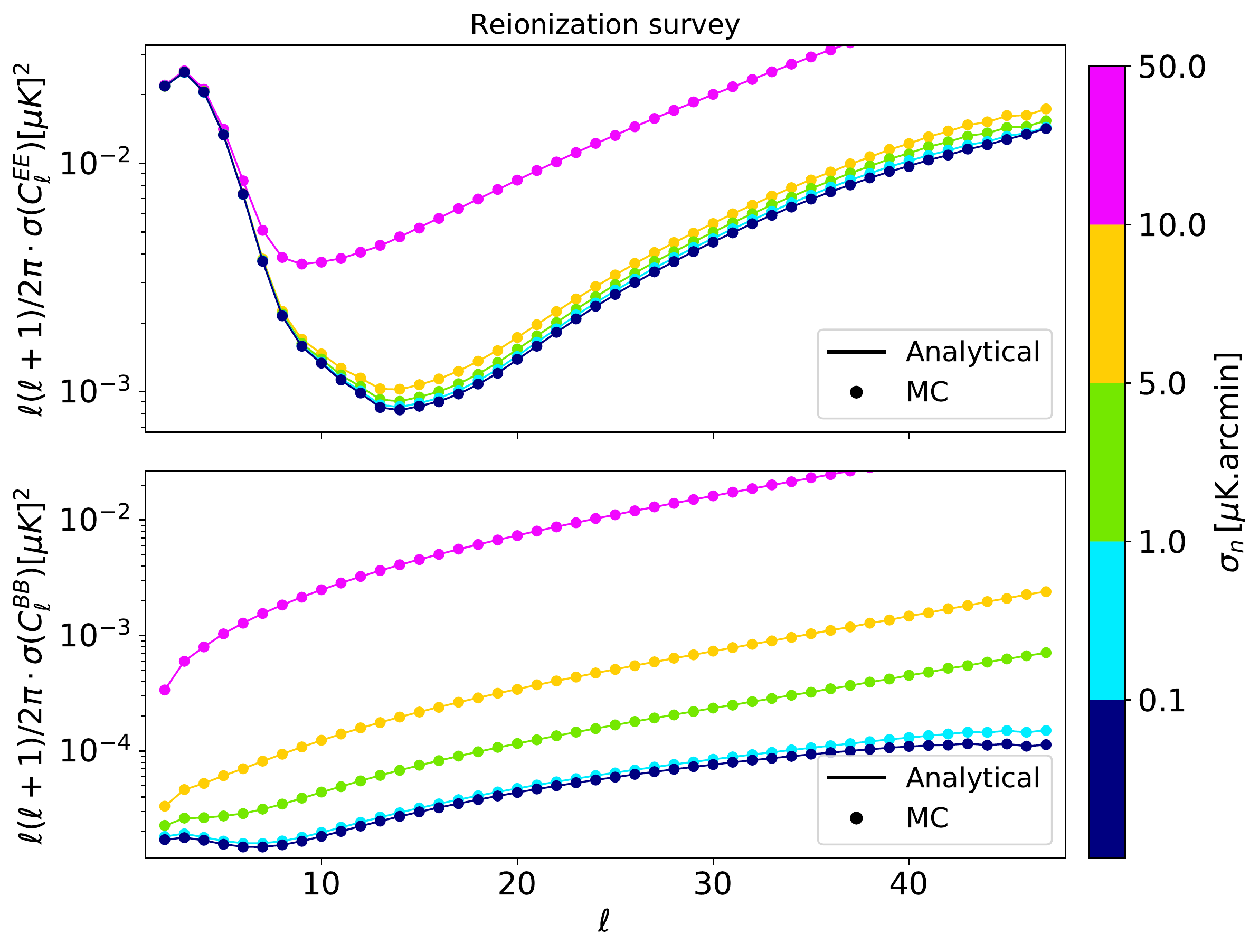}
\includegraphics[width=\columnwidth]{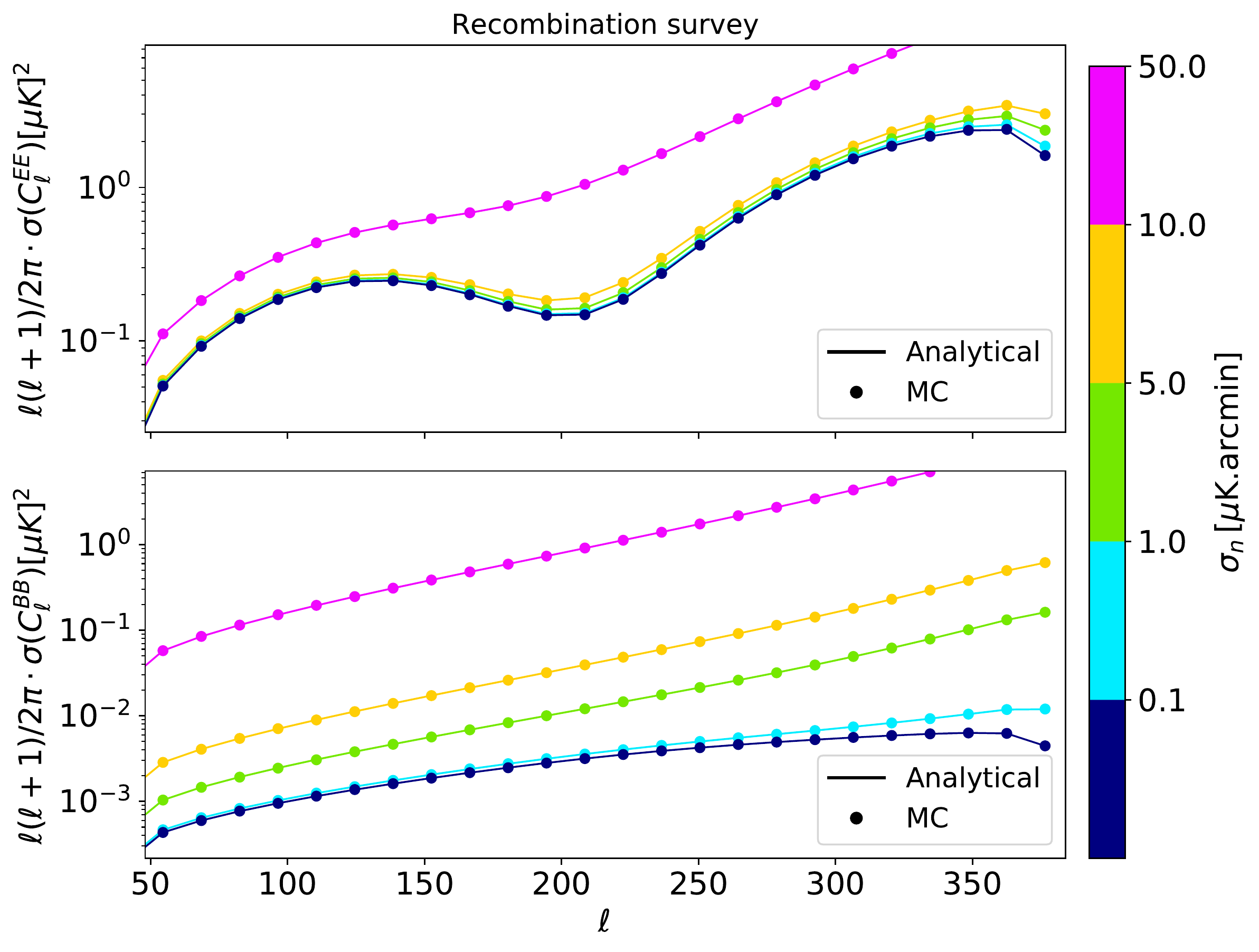}
\caption{Monte-Carlo (dots) and analytical (plain) errors of polarization spectra $EE$ (up) and $BB$ (bottom), for the reionization (left) and recombination (right) surveys, with noise levels $0.1\leq \sigma_n \leq50\,\mu \rm K .\rm arcmin$.\label{fig:Vars}}
\end{figure*}

\subsection{\label{sec:results}Power spectra reconstruction}

We verify with simulations that the reconstructed power spectra are unbiased with respect to the input model $C_\ell$. From the central limit theorem we expect that, as $n_{\rm MC}$ is large, the mean spectra residues
\begin{equation}
R_\ell[\hat C_\ell] \equiv \frac{C_\ell - \braket{\hat C_\ell}}{\sqrt{ \sigma^2(\hat C^{\rm MC}_\ell) / n_{\rm MC} }}\label{eq:ResidSpect}
\end{equation}
are expected to be normally distributed around zero, for all $\ell$ if the spectra are unbiased, with $\sigma^2(\hat C_\ell^{\rm MC})/ n_{\rm MC}$ the MC variance of the mean spectra. We carefully checked that this is the case for all noise levels $0.1\leq \sigma_n \leq50\,\mu \rm K . arcmin$. Power spectra and their residues are shown in Fig.~\ref{fig:ThvsMC} for $1\,\mu \rm K.arcmin$. Given the residues distribution for $n_{\rm MC}=10^{5}$ simulations, we conclude that the spectra bias level is less than one percent of the spectra errors.

The MC spectra variance, and that derived analytically $\sigma^2 (\hat C_\ell^{\rm ana}) = V_{\ell\ell}$ in Eq.~\eqref{eq:crossCovClEl} are shown to be in excellent agreement, as displayed in Fig.~\ref{fig:Vars}. The covariance matrix, not shown here, is band diagonal over the whole multipoles range, meaning that correlations are low and only occur between neighboring bins.

We successfully verified that the xQML spectrum reconstruction is unbiased, and that the MC covariance corresponds to that which is expressed analytically. The xQML thus gives a near minimal spectrum error.

%---------------------------------------------------------------------------------------------------------------------------------
% LEAKAGE
%---------------------------------------------------------------------------------------------------------------------------------

\section{\label{sec:Leakage}\texorpdfstring{$E$}-\texorpdfstring{$B$}{} leakage}

\subsection{\label{sec:ModeMix}Modes mixing}
\begin{figure}[!h]
\includegraphics[width=0.9\columnwidth]{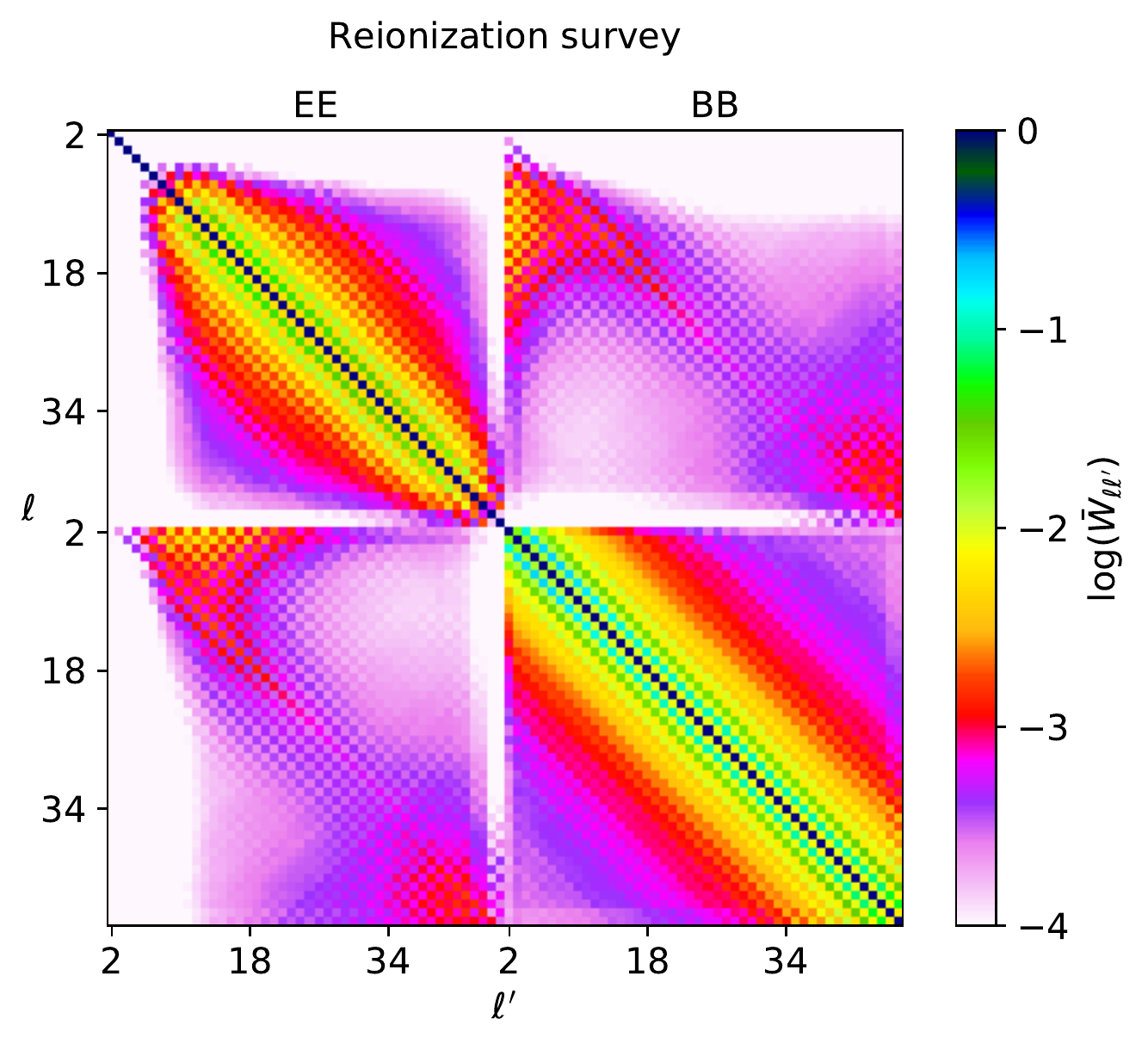}
\includegraphics[width=0.9\columnwidth]{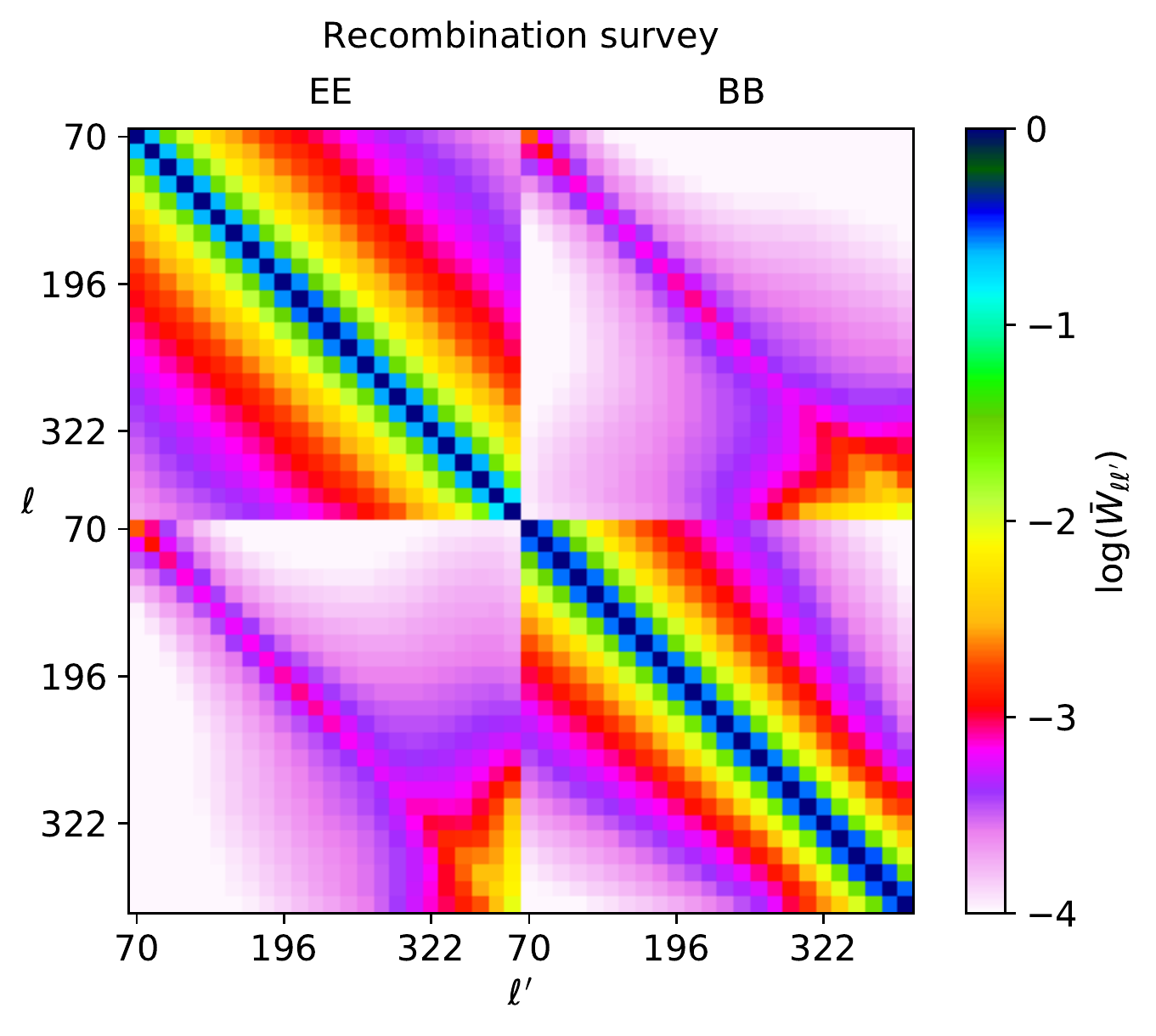}
\caption{The normalized mode-mixing matrix $\bar W_{\ell\ell'}$ defined in Eq.~\eqref{eq:CorrW} in log scale, for the reionization (up) and recombination (down) surveys, for $\sigma_n = 1\,\mu \rm K.arcmin$.\label{fig:WindowMat}}
\end{figure}

\begin{figure*}[!ht]
\includegraphics[width=0.49\textwidth]{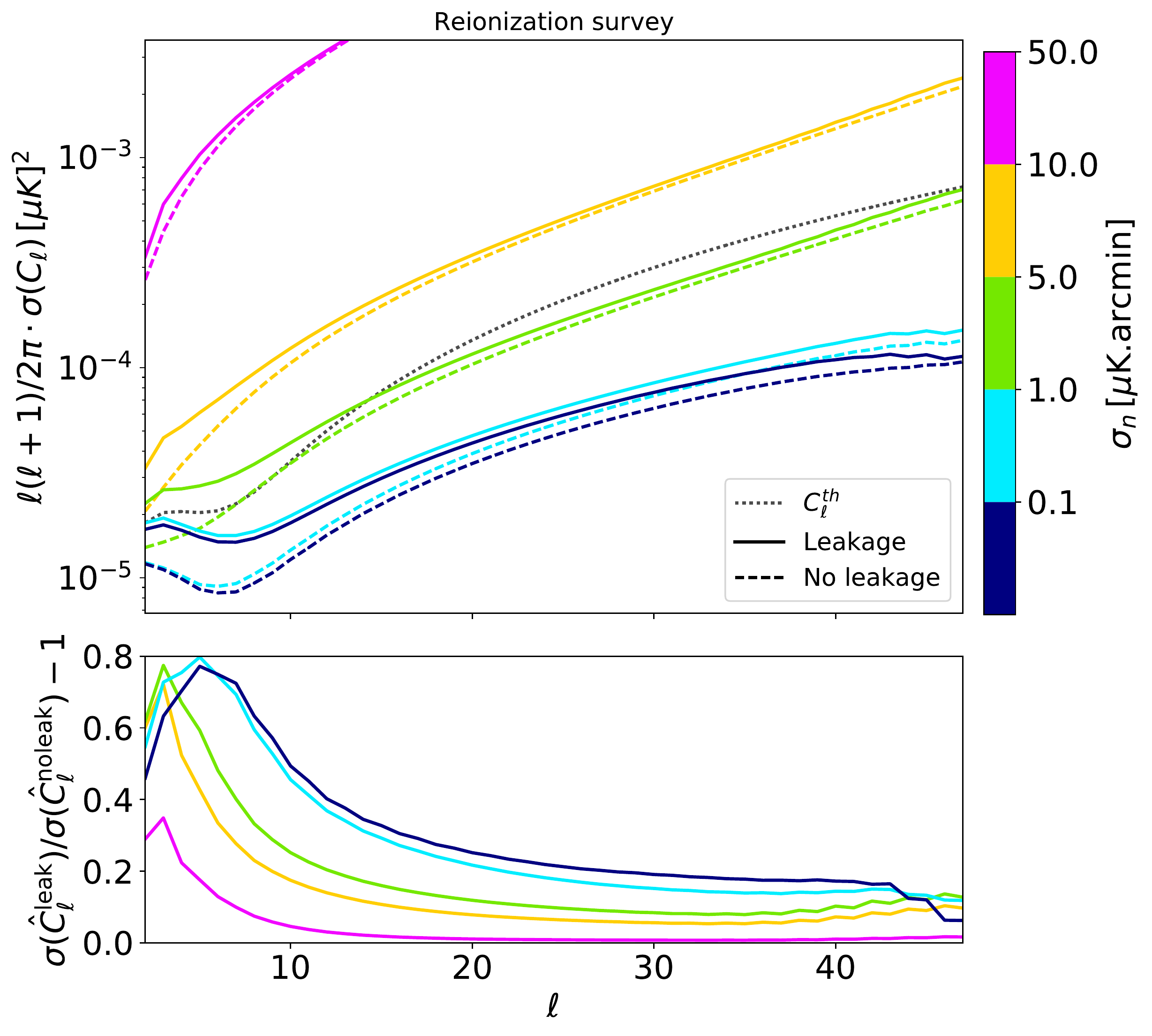}
\includegraphics[width=0.49\textwidth]{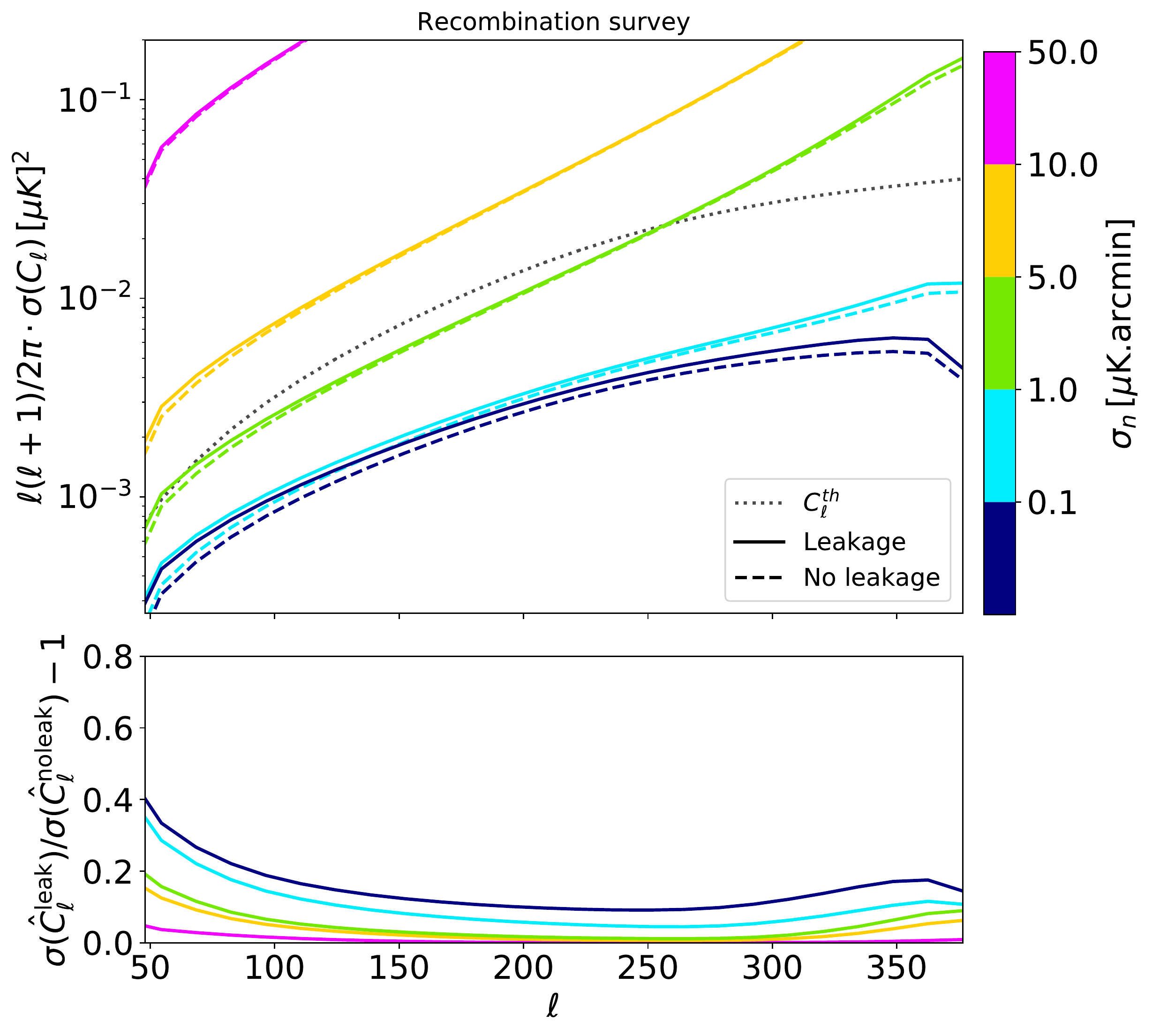}
\caption{Top panels show the $BB$-spectrum uncertainty with variance leakage (solid) and without (dashed), for the reionization (left) and recombination (right) surveys, at noise levels $0.1\leq \sigma_n \leq50\,\mu\rm K . arcmin$. 
Bottom panels quantify the absolute variance leakage, computed from $[\sigma (\hat C^{\rm leak}_\ell) - \sigma (\hat C^{\rm no leak}_\ell)]/\sigma(\hat C^{\rm no leak}_\ell)$.\label{fig:VarLeak}}
\end{figure*}

The mode-mixing matrix $W_{\ell\ell'}$ introduced in Eq.~\eqref{eq:autoWindow} quantifies the contribution of all $\ell'$-modes to the spectrum estimator at angular scale $\ell$. The rescaled matrix 
\begin{equation}
\bar W_{\ell\ell'} = \frac{W_{\ell\ell'}}{\sqrt{W_{\ell\ell}W_{\ell'\ell'}}},\label{eq:CorrW}
\end{equation}
is displayed in Fig.~\ref{fig:WindowMat} in log-scale for $\sigma_n = 1\,\mu \rm K.arcmin$. The off-diagonal blocks quantify the $E/B$ modes mixing, also known as polarization leakage. This mixing appears as soon as maps are partially masked, making some modes ambiguously belong to both $E$ and $B$ polarizations patterns.

We remark that the $E/B$ mixing is on average very low. Most of the $E$-to-$B$ leakage is localized at $\ell\lesssim 10$ for the reionization survey. The recombination survey also suffers from a polarization mixing increase at $\ell\gtrsim 250$. This effect is caused by the pixel resolution of the maps. It appears when the multipole angular scale is close to the typical pixel scale, and disappears as soon as we increase the datasets pixel resolution. The effect remains however very small. For the multipoles ranges of interest, it induces a negligible increase of variance as shown hereafter.

% Greetings, stranger. You found the easter egg :)

\subsection{\label{sec:VarianceLeakage}Variance induced leakage}

Because of polarization leakage, $E$ and $B$ modes respective uncertainty contribute to each other variance. For noise dominated datasets, this variance leakage has a small impact since both polarizations have the same noise, and their mutual contributions are equivalent. Conversely, when the noise is much below the signal level, the uncertainty is limited by the intrinsic 'cosmic variance', arising from the finite number of modes that can be sampled on the sky. The $E$-modes signal, thus its cosmic variance, is much higher than that of $B$-modes. As a consequence, even for small polarization mixing, the impact of the $E$-to-$B$ variance leakage can become non-negligible. 

Since, by construction the error of the xQML estimator is minimal, it also minimizes the amount of variance leakage. The $BB$ uncertainty is represented in Fig.~\ref{fig:VarLeak}, for which we compare the cases with and without leakage. The latter is obtained by simulating CMB polarization maps using null $EE$ and $TE$ spectra. We also show the absolute level of variance leakage $[\sigma (\hat C^{\rm leak}_\ell) - \sigma (\hat C^{\rm no leak}_\ell)]/\sigma(\hat C^{\rm no leak}_\ell)$. We observe that the recovered spectra uncertainties for $\sigma_n=0.1$ and $\sigma_n=1\,\mu\rm K.arcmin$ are both mostly cosmic variance limited by the lensing $B$-modes signal. We also recover that the impact of the variance leakage gets less important as the SNR decreases.

For the reionization survey, the variance leakage is observed to be maximal at large angular scales, up to a $80\%$ increased uncertainty around $\ell \lesssim 10$, which quickly drops to $30\%$ for higher $\ell$'s. This is not surprising since, for this multipole's range, the $EE$ cosmic variance as well as the $E$-to-$B$ mixing in $\bar W_{\ell\ell'}$ are maximal.

For the recombination survey, the impact is maximal for the first bins. This is again related to the higher polarization mixing in $\bar W_{\ell\ell'}$ at those multipoles. It then drops to $20\%$ for $\ell\gtrsim90$, followed by a slight increase at $\ell\gtrsim250$. This is consistent with the previous $E/B$ mixing observations made on $\bar W_{\ell\ell'}$ for this multipoles range. The impact at low $\ell$'s remains however smaller since the $E$-modes cosmic variance level is much lower for those angular scales. 

We conclude that, even if the mixing between polarization modes is minimized when using the xQML estimator, the induced variance increase can however be non-negligible, especially at large angular scales. 

\subsection{\label{sec:CompPpCl}Comparison with pseudo-spectra}

\begin{figure*}[!ht]
		\includegraphics[width=0.49\textwidth]{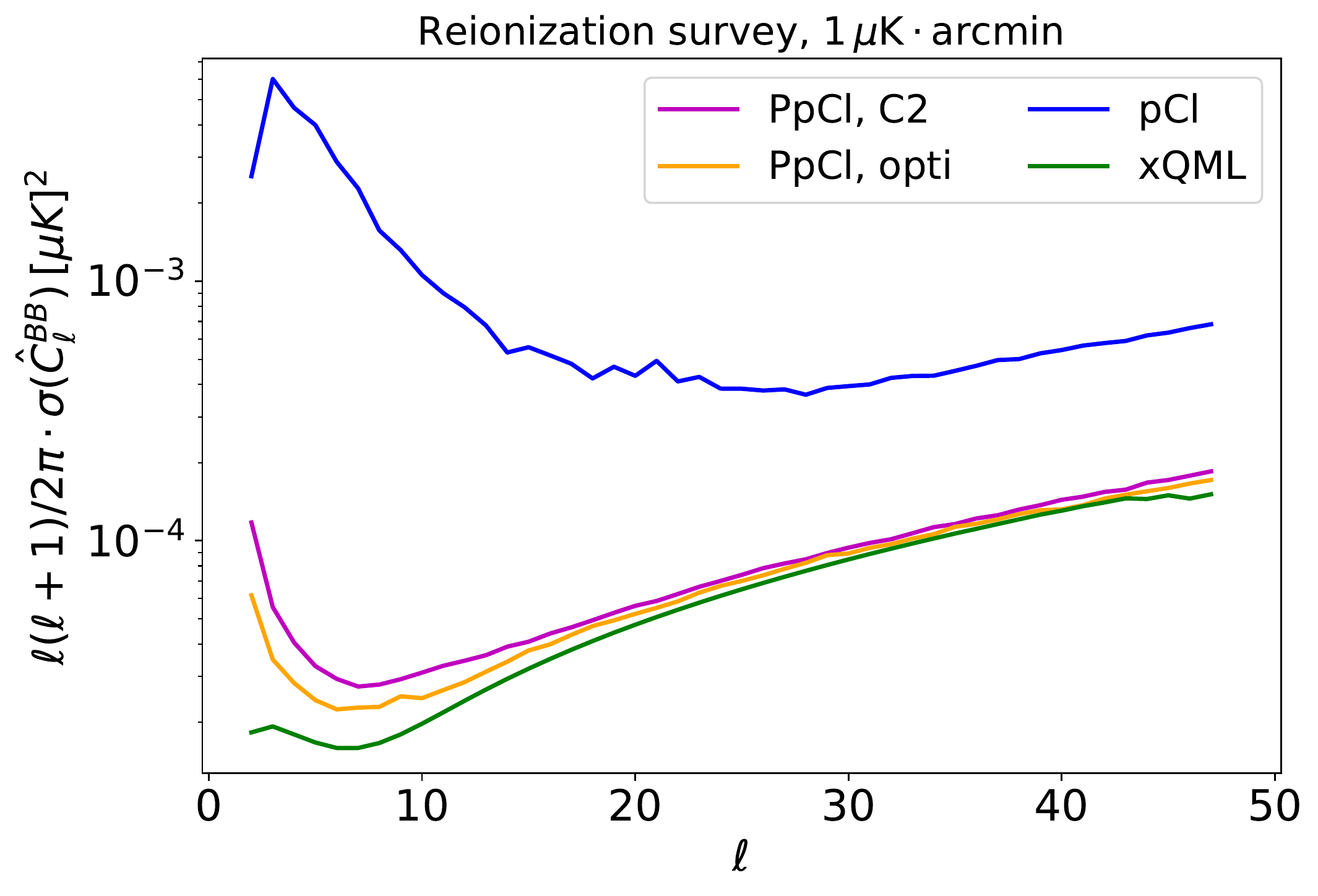}
		\includegraphics[width=0.49\textwidth]{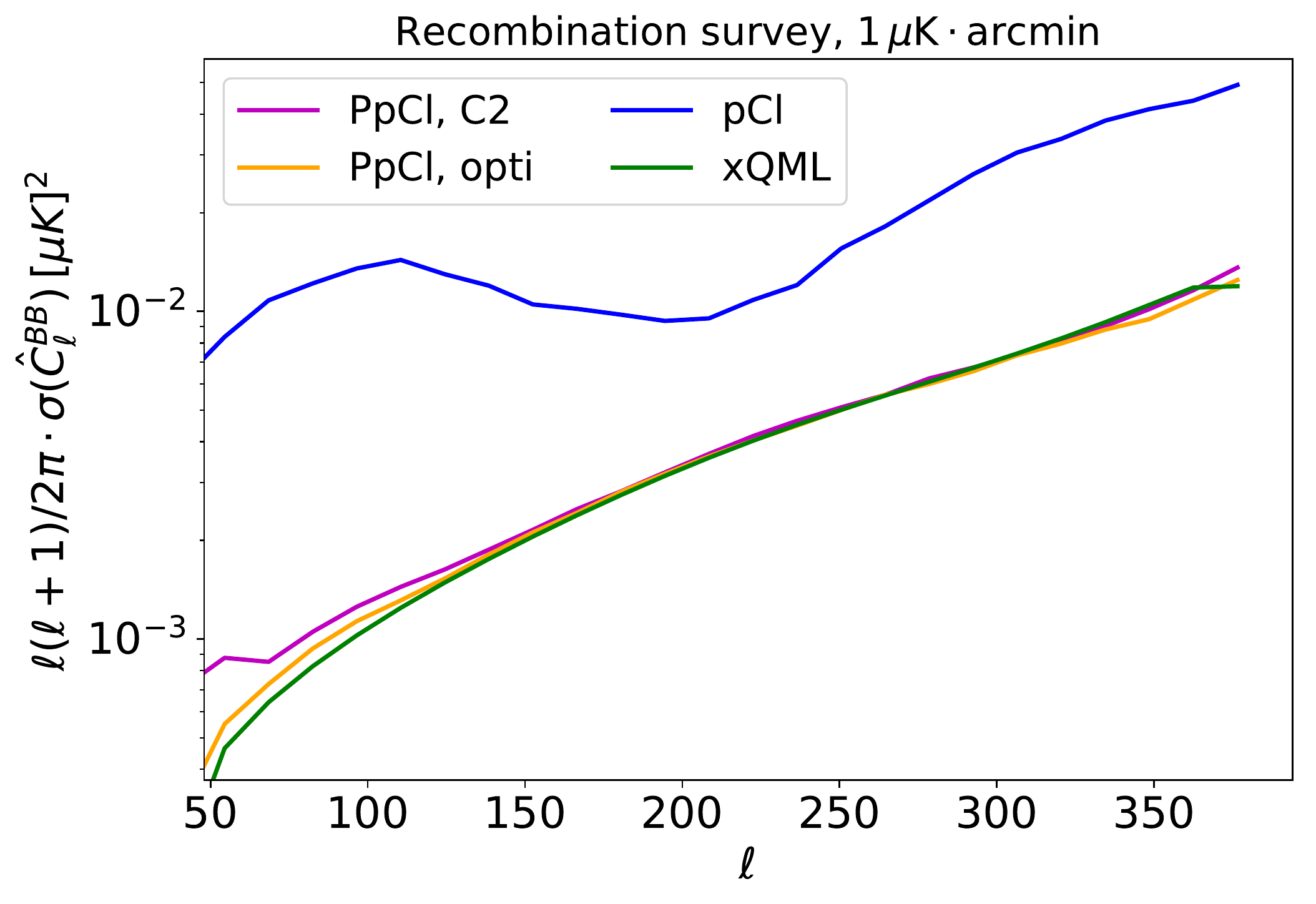}
\caption{$BB$ spectrum errors from xQML, standard pCl estimators, PpCl (pure pCl) using the 'C2' and optimized apodization, for the reionization (left) and recombination (right) surveys, with a noise level $\sigma_n = 1\,\mu{\rm K . arcmin}$.}
\label{fig:PureVarBB}
\end{figure*}

In this section we compare the xQML with other methods such as the standard pCl, and the (pure) PpCl approach. The latter method requires the mask and its first derivative to be equal to zero on its boundaries in order to eliminate the polarization variance leakage. We follow the cross PpCl formalism described in~\cite{grain_polarized_2009}, using the two mask apodization processes. 

The first process is achieved by isotropically applying the apodization function 'C2' defined in~\cite{grain_polarized_2009} and parametrized by the apodization length parameter $\theta_*\,[\rm deg]$, to our binary masks. This parameter needs to be adapted to the SNR. We thus select for each multipole the mask apodization for which the mode estimate has minimal variance. Each mode is then combined to reconstruct an unbiased spectrum estimation, which the covariance matrix can be evaluated using MC simulations. This process has to be repeated depending on the noise level. 

An optimized apodization process, proposed in \cite{smith_general_2007}, consists of finding the adequate windows functions that lower the total (noise and leakage) $B$ pseudo-spectrum variance. This is achieved by implementing an iterative Preconditioned Conjugate Gradient solver as in ~\cite{smith_general_2007, grain_polarized_2009}. In this framework, the relation between the mask and its derivative required by the pure method is relaxed. The optimization is performed over bins of multipoles. 
The mixing kernel that allows us to recover the spectra from the pseudo-spectra is then built according to each bin optimization.

The uncertainty on the reconstructed $B$-mode power spectrum for methods introduces above are illustrated in Fig.~\ref{fig:PureVarBB} based on $10^4$ MC simulations. The standard pCl leads to higher uncertainties, for which the $E$-modes variance leakage contribution is particularly visible on the recombination survey $B$-modes variance. 

The pure methods PpCl allow to recover much lower error bars. The two apodizations gives similar results at high multipoles but an optimized apodization is required to obtain better results at large angular scales.
Nevertheless, the pixel domain cross-correlation xQML provides the lowest spectra uncertainty over the whole multipole range. 
This is particularly true at large angular scales, and even increases when the tensor-to-scalar ratio $r$ decreases.

We conclude that the xQML method is particularly suited for reducing the $B$-modes variance leakage for large angular scale analysis compared to the PpCl approach. It produces smaller error bars and does not require mask apodization optimizations. This is of special interest for the detection of primordial $B$-modes.

\section{\label{sec:EBspec}\texorpdfstring{$E$-$B$}{} correlation spectrum}

Although first order primordial $E$-$B$ and $T$-$B$ correlations are predicted to be null in the frame of the $\Lambda\rm CDM$ model, nonstandard cosmological mechanisms, such as cosmic birefringence, could induce nonzero correlation spectra~\cite{lue_cosmological_1999,carroll_quintessence_1998,loeb_faraday_1996,kahniashvili_effects_2005,campanelli_faraday_2004,caprini_cosmic_2004,pogosian_signatures_2002}. In addition to providing an important probe to nonstandard physics, measuring $EB,TB$ spectra could also help to diagnose instrumental systematic effects~\cite{yadav_primordial_2010,hu_benchmark_2003}.

\begin{figure}[!htb]
\includegraphics[width=\columnwidth]{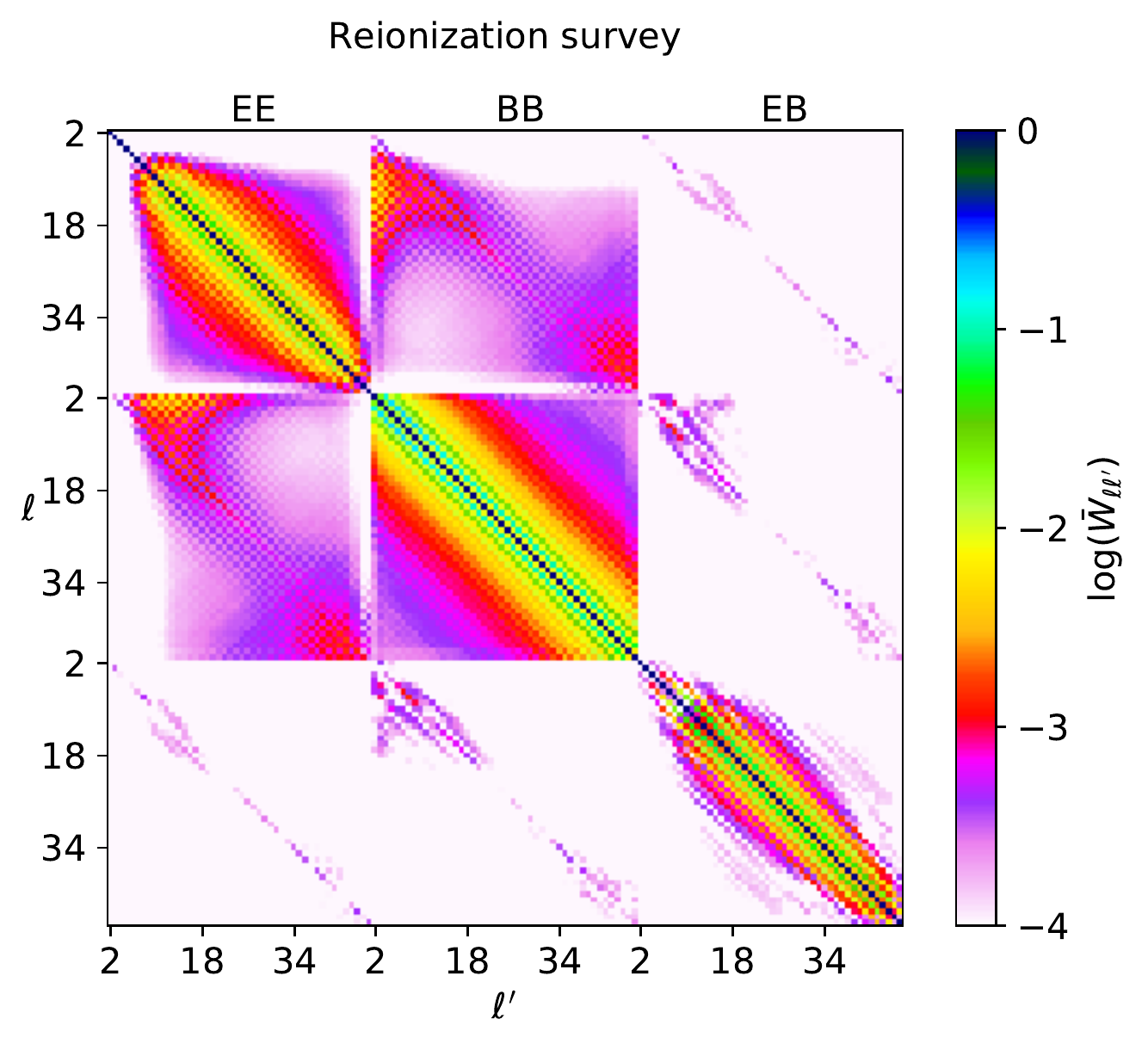}\\~\\
\includegraphics[width=\columnwidth]{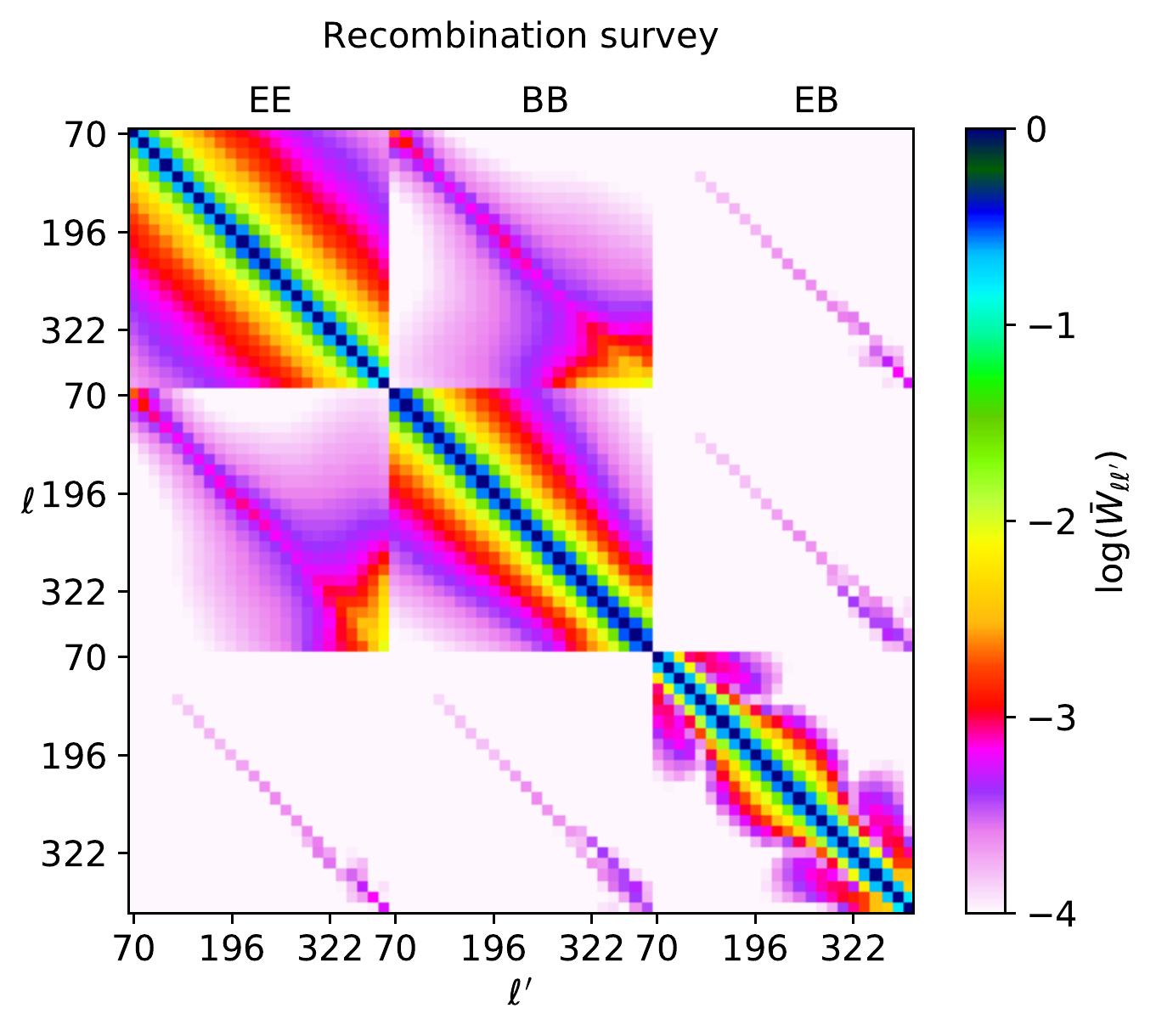}
\caption{The normalized mode-mixing matrix $\bar W_{\ell\ell'}$ defined in Eq.~\eqref{eq:CorrW} in log scale, for the reionization (up) and recombination (down) surveys, for $\sigma_n = 1\,\mu \rm K.arcmin$.\label{fig:WindowMatEB}}
\end{figure}

We focus on the $E$-$B$ correlation, for which we compute the $EB+BE$ spectrum variance. The rescaled mode-mixing matrix introduced in Eq.~\eqref{eq:CorrW} is extended to $EB$ multipoles as displayed in Fig.~\ref{fig:WindowMatEB} for $1\,\mu\rm K.arcmin$. Apart from a negligible resolution effect for high $\ell'$'s, we observe no mixing between $EB$ and $EE,BB$ when using the xQML method. Note, however that this statement is not true if we consider particular models with nonzero $\tilde C^{EB}_\ell$.

\begin{figure*}[!ht]
		\includegraphics[width=0.49\textwidth]{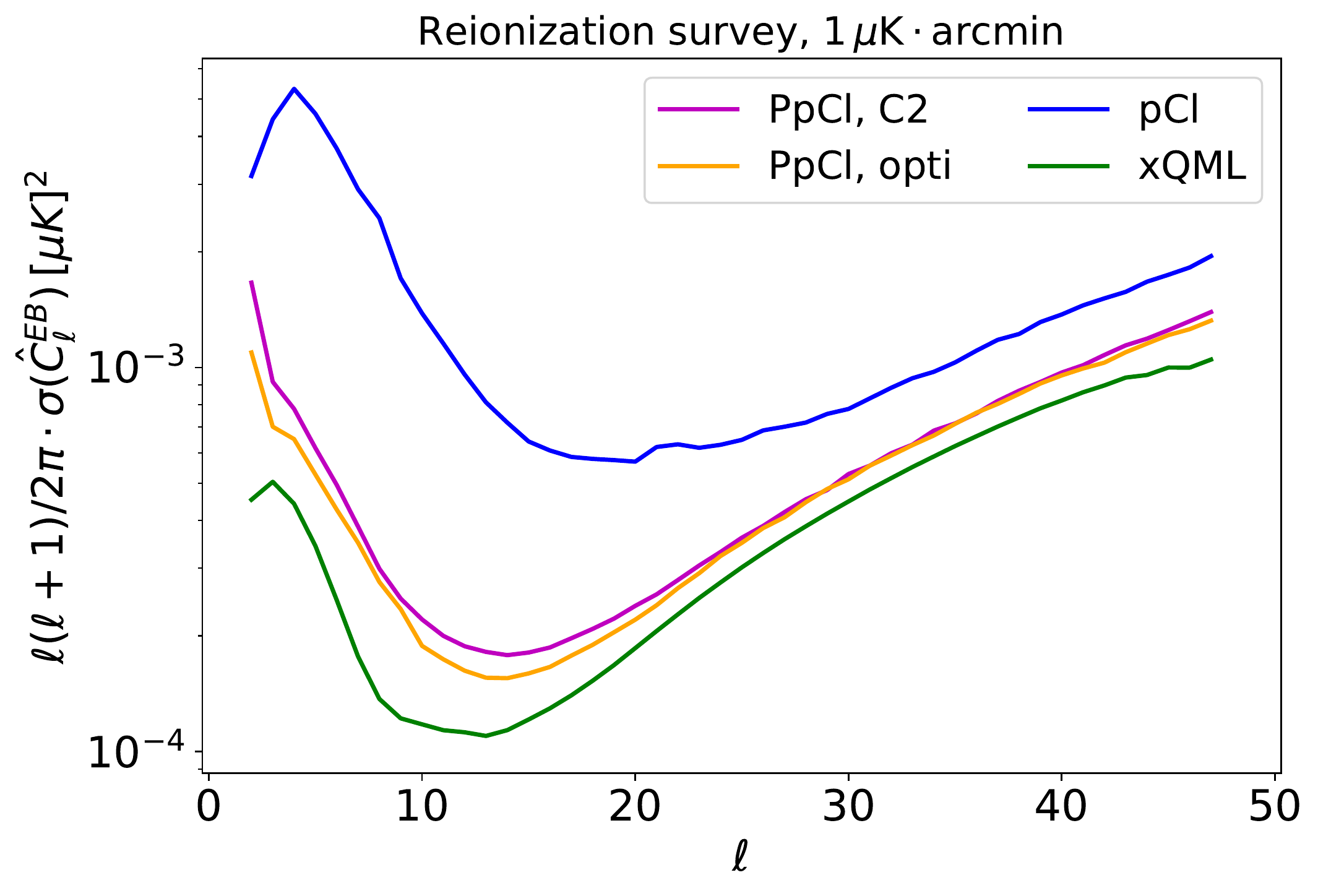}
		\includegraphics[width=0.49\textwidth]{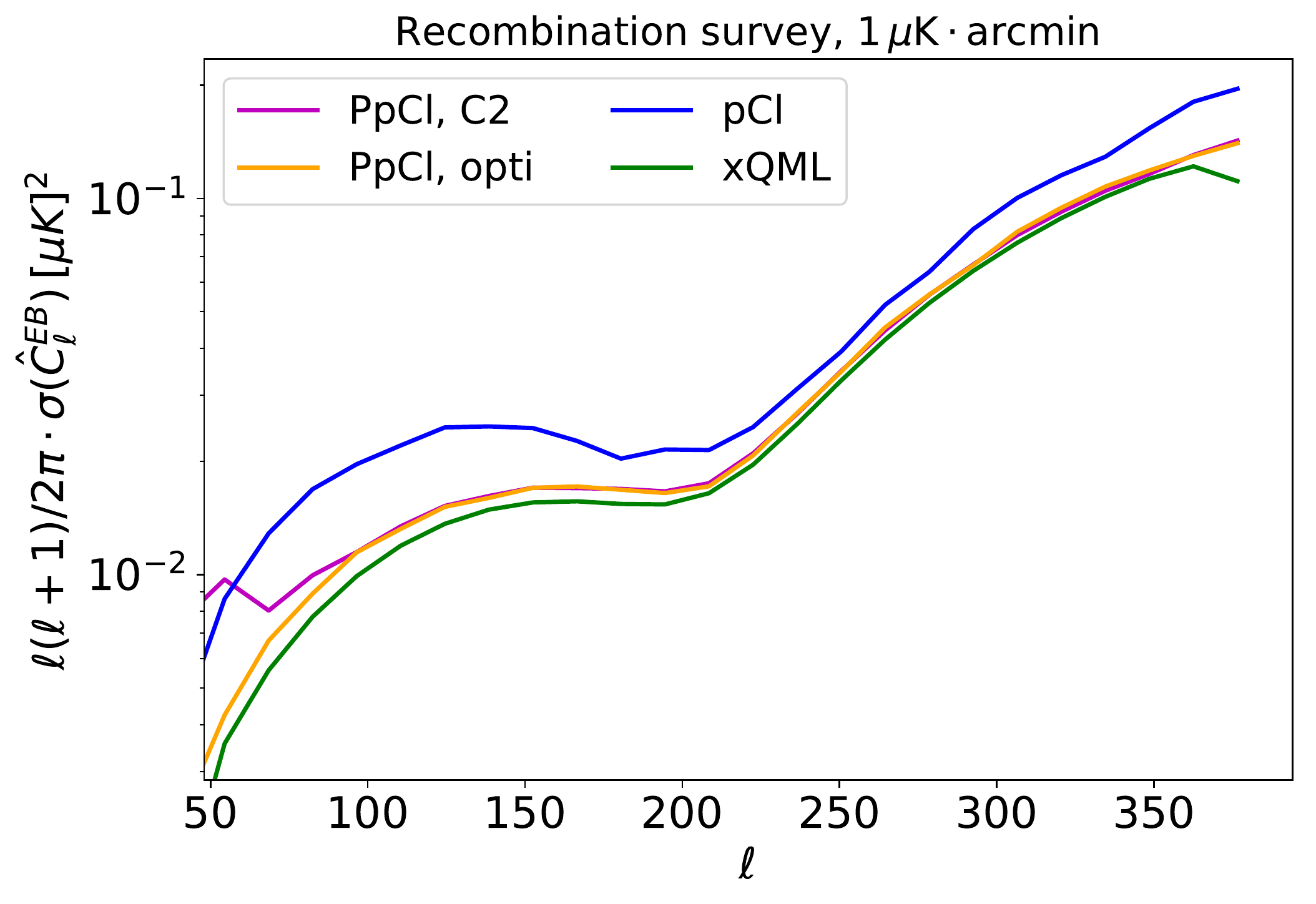}
\caption{$EB$ spectrum errors from xQML, standard pCl estimators, and PpCl (pure pCl) using the 'C2' and optimized apodization, for the reionization (left) and recombination (right) surveys, with a noise level $\sigma_n = 1\,\mu \rm K . arcmin$.}
\label{fig:PureVarEB}
\end{figure*}

As in the previous section for the $BB$ uncertainty, we compare our results with the pCl and PpCl methods. The latter is computed using the hybrid approach proposed in~\cite{grain_cmb_2012}, where the $E$-modes are obtained using the standard pseudo-spectrum, and the $B$-modes using the pure method. Variances are shown in Fig.~\ref{fig:PureVarEB} for $1\,\mu\rm K.arcmin$. 
The PpCl uncertainty is about $20\%$-$60\%$ higher than that of the xQML for the reionization survey. Longer mask apodization lengths improve the PpCl error for $\ell\lesssim 10$.
On the recombination survey, the xQML gives significant lower $EB$ uncertainty only for $\ell\lesssim 100$.
The conclusion is similar as for the $BB$-spectrum analysis. The xQML method provides an efficient estimator for large angular scales analysis.

%---------------------------------------------------------------------------------------------------------------------------------
% CONCLUSION
%---------------------------------------------------------------------------------------------------------------------------------

\section{\label{sec:Conclusion}Conclusion}

In this paper, we derived a pixel-based spectrum estimator that allows us to cross-correlate CMB datasets. The method is very similar to the QML, but does not require a precise knowledge of the datasets noise covariance matrices to subtract the noise bias. We also provided an approximation to the Sylvester equation that has little impact on the optimality of the estimator, which, by construction, provides near-minimal error bars. The estimator variance is shown to be sensitive to only second order perturbations of the fiducial pixels covariance matrix. Moreover, using no $TQ$ and $TU$ correlations for the construction of this matrix, temperature and polarization analysis can be done completely separately.

We showed that the xQML estimator is unbiased, and that the error bars on the recovered spectrum, obtained from Monte-Carlo simulations, correspond to the analytically derived variance. We presented two CMB surveys aiming at the reionization and recombination polarized signals measurement, with a fiducial tensor-to-scalar ratio $r = 10^{-3}$. The source of polarization leakage can be identified in the mode-mixing matrix $W_{\ell\ell'}$. We showed in Sec.~\ref{sec:Leakage} that it is consistent with the increase of variance in $B$-modes when compared to the no-leakage case. The reionization survey $BB$ uncertainty at low noise levels is particularly impacted by the polarization mixing, with a maximum of an $80\%$ increase for large angular scales at $0.1$ - $1\,\mu\rm K.arcmin$. Since the xQML method minimizes bins correlations as well as polarization mixing, the resulting error bars thus correspond to the minimal uncertainty achievable when aiming to polarization variance leakage reduction.

Comparison with the pure pseudo-spectrum formalism shows a significant improvement of the error bars and correlations for both $BB$ and $EB$ when using the xQML method. The particular advantage relative to pure methods is that it does not require any special masks apodization processing. However, due to its higher computational cost ($\mathcal O (N_{\rm d}^3)$ operations) relative to pseudo-spectra ($O (N_{\rm d}^{3/2})$ operations), the xQML cannot be run on as many multipoles as for the pseudo-spectra. For those reasons, the xQML estimator is particularly suited for large and intermediate angular scales analysis.

\begin{figure}[!htb]
\includegraphics[width=\columnwidth]{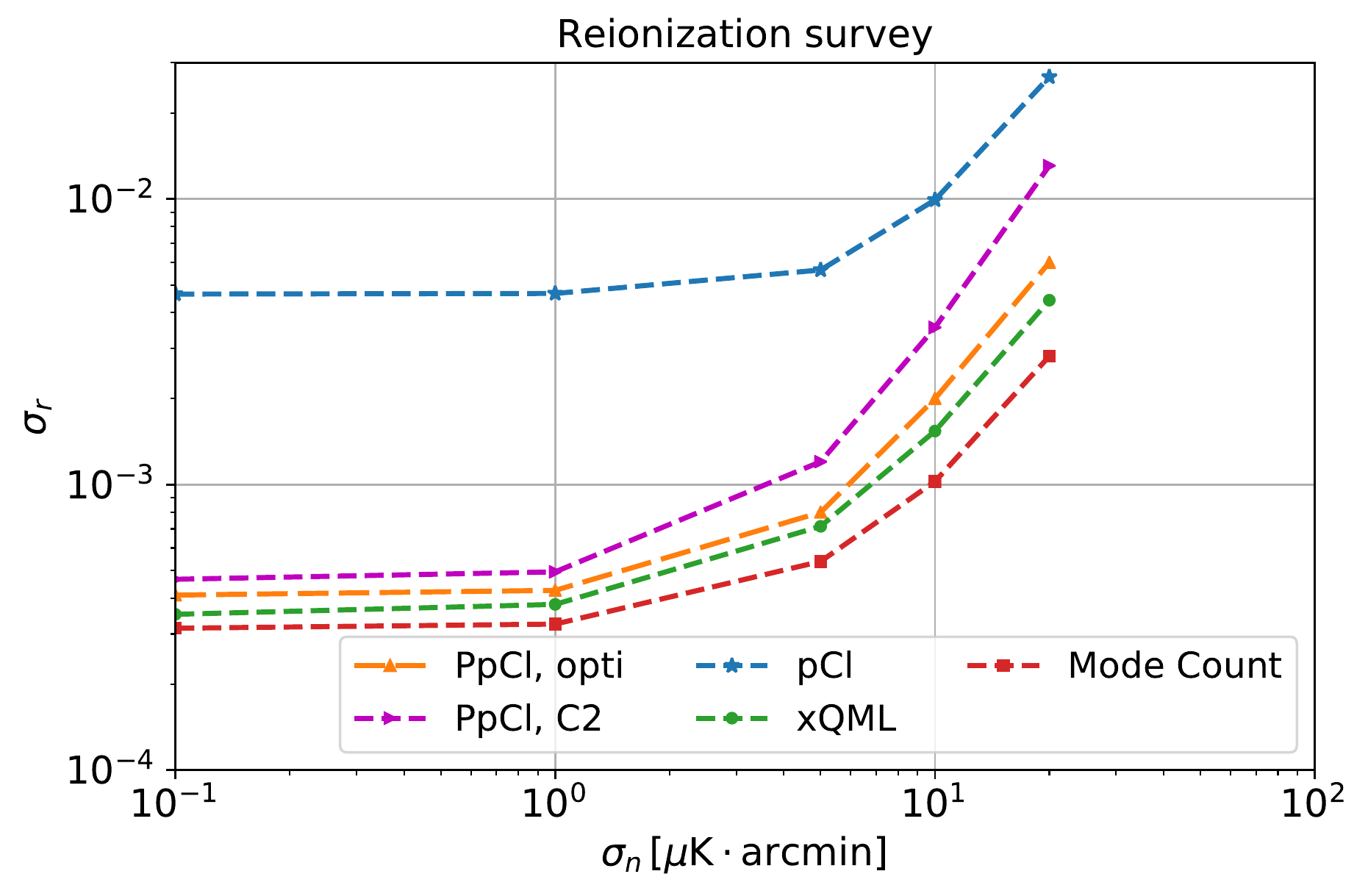}
\includegraphics[width=\columnwidth]{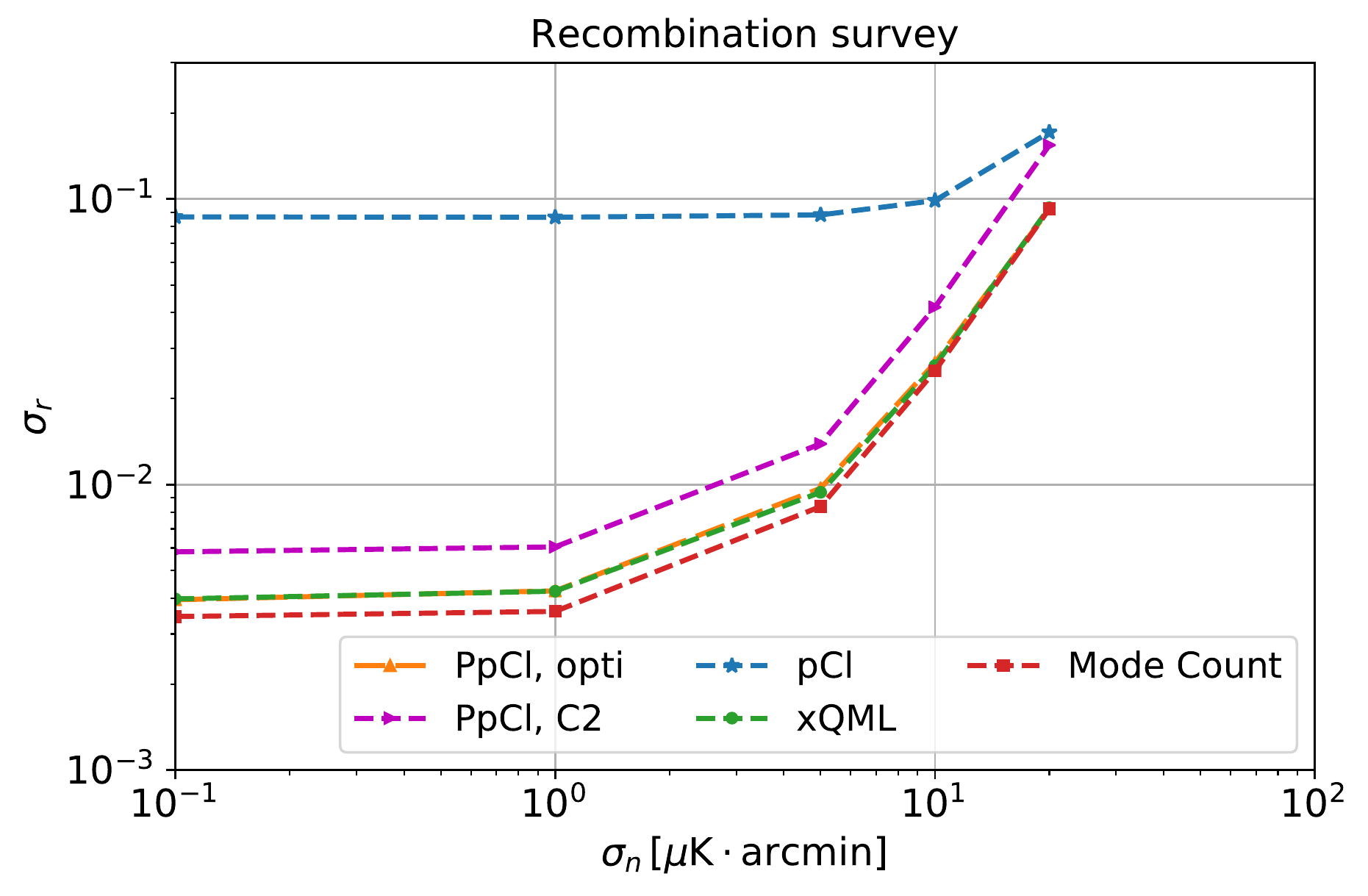}
\caption{Error on the tensor-to-scalar ratio with a fiducial $ r=10^{-3}$, for the reionization (up) and recombination (bottom) surveys as a function of the noise levels $0.1 \leq \sigma_n \leq 20\,\mu\rm K.arcmin$. We compare the uncertainty obtained from the standard pseudo-spectrum (blue), the pure pseudo-spectrum using C2 (magenta) and optimized apodization (orange), the cross quadratic pixel based (green), and the mode-counting formula (red).}\label{fig:sigR}
\end{figure}

As a forecast analysis, we show in Fig.~\ref{fig:sigR} the uncertainty of $r$, obtained from each method introduced previously, as a function of the noise level. We also proceed to a comparison with the mode-counting formula\footnote{I.e. $\sigma^2_{\rm m.c.} =  \left[ 2  \tilde C_\ell^2 + \tilde C_\ell(N_\ell^{A} + N_\ell^{B}) + N_\ell^{A} N_\ell^{B}   \right]/\left[  \ell(\ell+1) \Delta_\ell f_{\rm sky} \right] $, where $\tilde C_\ell$ is the power spectrum fiducial model, $N_\ell = n_\ell/B_\ell^2$ is the noise spectrum of the dataset convolved by the corresponding beam functions $B_\ell$.
}, which gives a naive estimate of the lowest achievable variance, neglecting correlations and leakage induced by the sky coverage. We use the spectrum-based likelihood presented in~\cite{mangilli_large-scale_2015}, which is a cross-spectra extended version of the low-multipoles Hamimeche and Lewis likelihood~\cite{hamimeche_likelihood_2008}. The pure method covariance matrix is computed using MC as described in Sec.~\ref{sec:CompPpCl}. We consider only two datasets, no foreground contamination and/or residuals, nor de-lensing. For low SNR, the impact of the polarization mixing is small, and all (standard and pure) pseudo-spectrum methods give the same error on $r$. For high SNR, the uncertainty of $r$ is cosmic variance limited, which corresponds to the plateau from $\sigma_n=0.1$ to $\sigma_n=1\,\mu\rm  K . arcmin$. In this range of noise level, the pure pseudo-spectrum method with optimized apodization and the xQML gives the same uncertainty of $r$ for the recombination survey, while the xQML uncertainty is $\sim20\%$ lower than the optimized PpCl method for the reionization survey.

\begin{acknowledgments}
The authors would like to thank G. Efstathiou and J. Grain for the helpful discussions about the xQML and pure methods. We also thank H. Taeter for his help leading to the identification of the Sylvester equation. Some of the results in this paper have been derived using the HEALPix \cite{healpix:_2018} package, the NaMaster~\cite{alonso_unified_2018} package, and the Xpure package~\cite{trsitram_xpure:_2018}.
\end{acknowledgments}

\clearpage
\bibliographystyle{apsrev}

\bibliography{xqmlref}

\end{document}